\documentclass[acmsmall]{acmart}

\acmConference
\acmBooktitle
\AtBeginDocument{%
  }

\setcopyright{acmlicensed}
\acmJournal{PACMHCI}
\acmYear{2025} \acmVolume{9} \acmNumber{2}
\acmArticle{CSCW142} \acmMonth{4}
\acmDOI{10.1145/3711040}

\usepackage{hyperref}

\usepackage{graphicx}

\usepackage{tabularx}
\usepackage{booktabs}
\usepackage{xcolor}

\usepackage{array}
\usepackage{colortbl}
\usepackage{multirow}
\usepackage{hhline}
\usepackage{tabularray}

\usepackage{arydshln}
\begin{document}

\title[Avatar Animation and Meeting Outcome]{Nods of Agreement: Webcam-Driven Avatars Improve Meeting Outcomes and Avatar Satisfaction Over Audio-Driven or Static Avatars in All-Avatar Work Videoconferencing}

\author{Fang Ma}
\email{f.ma@gold.ac.uk}
\orcid{0000-0001-5476-6982}
\affiliation{%
  \institution{Goldsmiths, University of London}
  \city{London}
  \country{United Kingdom}
}

\author{Ju Zhang}
\email{j.zhang@gold.ac.uk}
\orcid{0009-0004-3655-4010}
\affiliation{%
  \institution{Goldsmiths, University of London}
  \city{London}
  \country{United Kingdom}
}

\author{Lev Tankelevitch}
\email{lev.tankelevitch@microsoft.com}
\orcid{0000-0003-1286-5194}
\affiliation{%
  \institution{Microsoft Research}
  \city{Cambridge}
  \country{United Kingdom}
}

\author{Payod Panda}
\email{payod.panda@microsoft.com }
\orcid{0000-0002-8208-4185}
\affiliation{%
  \institution{Microsoft Research}
  \city{Cambridge}
  \country{United Kingdom}
}

\author{Torang Asadi}
\email{torang.asadi@microsoft.com}
\orcid{0000-0003-2796-7230}
\affiliation{%
  \institution{Microsoft}
  \city{San Francisco}
  \country{United States}
}

\author{Charlie Hewitt}
\email{chewitt@microsoft.com}
\orcid{0000-0003-3943-6015}
\affiliation{%
  \institution{Microsoft}
  \city{Cambridge}
  \country{United Kingdom}
}

\author{Lohit Petikam}
\email{lohitpetikam@microsoft.com}
\orcid{0000-0001-6629-7490}
\affiliation{%
  \institution{Microsoft}
  \city{Cambridge}
  \country{United Kingdom}
}

\author{James Clemoes}
\email{jaclemoe@microsoft.com}
\orcid{0009-0009-5083-201X}
\affiliation{%
  \institution{Microsoft Research}
  \city{Cambridge}
  \country{United Kingdom}
}

\author{Marco Gillies}
\email{m.gillies@gold.ac.uk}
\orcid{0000-0002-3100-9230}
\affiliation{%
  \institution{Goldsmiths, University of London}
  \city{London}
  \country{United Kingdom}
}

\author{Xueni Pan}
\email{x.pan@gold.ac.uk}
\orcid{0000-0002-1282-1469}
\affiliation{%
  \institution{Goldsmiths, University of London}
  \city{London}
  \country{United Kingdom}
}

\author{Sean Rintel}
\email{serintel@microsoft.com}
\orcid{0000-0003-0840-0546}
\affiliation{%
  \institution{Microsoft Research}
  \city{Cambridge}
  \country{United Kingdom}
}

\author{Marta Wilczkowiak}
\email{mawilczk@microsoft.com}
\orcid{0009-0006-7695-4216}
\affiliation{%
  \institution{Microsoft}
  \city{Cambridge}
  \country{United Kingdom}
}

\renewcommand{\shortauthors}{Ma et al.}

\begin{abstract}
Avatars are edging into mainstream videoconferencing, but evaluation of how avatar animation modalities contribute to work meeting outcomes has been limited. We report a within-group videoconferencing experiment in which 68 employees of a global technology company, in 16 groups, used the same stylized avatars in three modalities (static picture, audio-animation, and webcam-animation) to complete collaborative decision-making tasks. Quantitatively, for meeting outcomes, webcam-animated avatars improved meeting effectiveness over the picture modality and were also reported to be more comfortable and inclusive than both other modalities. In terms of avatar satisfaction, there was a similar preference for webcam animation as compared to both other modalities. Our qualitative analysis shows participants expressing a preference for the holistic motion of webcam animation, and that meaningful movement outweighs realism for meeting outcomes, as evidenced through a systematic overview of ten thematic factors. We discuss implications for research and commercial deployment and conclude that webcam-animated avatars are a plausible alternative to video in work meetings.
\end{abstract}

\begin{teaserfigure}
  \centering
  \includegraphics[width=0.95\textwidth]{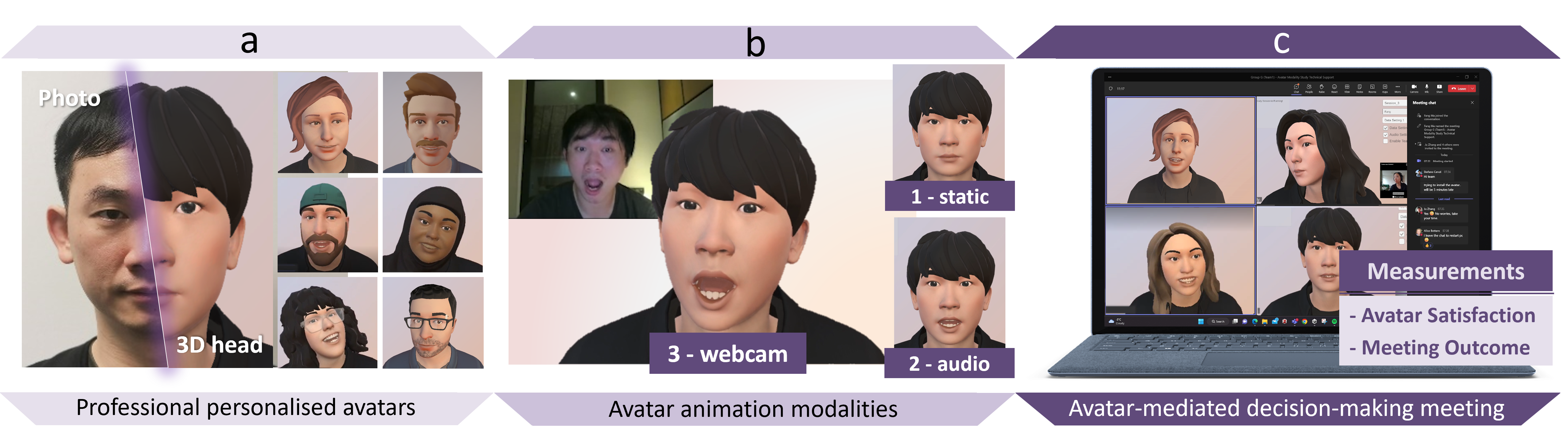}
  \caption{An overview of the experimental study. a) A personalized 3D stylized avatar was generated for each participant, using their photos with facial landmarks, and preserving diverse personal features like beards, glasses, and hairstyles. b) Participants used their avatars in 3 modalities: static picture, audio-animated and webcam-animated, affecting their head movements and expressions (see supplementary video). c) Participants used the avatars in a videoconferencing meeting, during which they engaged in three group decision-making tasks, one for each avatar modality. Meeting outcomes and avatar satisfaction were measured after each task and compared.}
  \Description{A Microsoft Teams meeting using experimental stylized avatars.}
  \label{fig:teaser}
\end{teaserfigure}



\received{January 2024}
\received[revised]{July 2024}
\received[accepted]{October 2024}

\maketitle

\section{Introduction}

In 2023, both Microsoft Teams \cite{MicrosoftTeams_avatars_2023} and Zoom \cite{zoom_avatars_2023} introduced stylised avatars for all users as an alternative to webcam video in their videoconferencing services. The ability to use avatars is a major change in representational affordances for users, especially in a work context, offering a crucial new choice for people who prefer not to use video or are unable to use video. From a commercial perspective, completely static avatars are obviously the cheapest to develop and use, albeit likely to have low user satisfaction. The currently available avatars in Zoom and Microsoft Teams are limited to viseme animation (audio-driven mouth positions) and some pre-designed animations, which limit the naturalness and fidelity of their expression compared with actual video. User satisfaction and meeting outcomes are likely to improve with more signals for animating avatars, but increased expressiveness is also associated with increased effort and time to develop at production quality. On the other hand, minimal animation also offers benefits, such as protecting user privacy and reducing bandwidth consumption in virtual meetings~\cite{panda_avatarmeetingalltogether_2022, abdullah_videoconference_2021_[b]}. Navigating the trade-off between increased expressiveness and the required development effort remains a challenge.

Previous studies have explored various aspects of avatar-mediated communication. 
While older studies supported the notion that stylized avatars are inappropriate for professional settings~\cite{Junuzovic2012ToSee}, recent studies of avatars in workplace and classroom settings show that this trend might be changing~\cite{BusinessMeetings, Laboratory}. 
One key result from recent avatar studies is that the visual fidelity of avatars is less relevant than their motion fidelity, which holds for both self-identification~\cite{mar_enfacement_motion_2020} and trusting other avatar participants in a meeting~\cite{panda_avatarmeetingalltogether_2022}. 
However, the link between avatar motion fidelity and work meeting outcomes remains unclear. More broadly, avatar research on meeting outcomes is very limited \cite{dobre_2022}.
Critically, both the scientific and commercially-driven questions are the same: 
\textbf{what are the differences between avatar animation modalities in terms of their impact on meeting outcomes and user preferences?}

To answer these research questions, we conducted a within-subjects, mixed-methods experiment in which 68 employees of a global technology company (in 16 groups), represented as stylized avatars (with a professional appearance), engaged in collaborative decision-making tasks akin to those in work meetings. The avatars all used the same illustration style, but had three modalities of animation, randomised between tasks in a counterbalanced order across groups: no animation (picture), audio-animation (visemes), and webcam-animation (visemes, facial expressions, and head movements).

Our quantitative results demonstrate that webcam-animated improves meeting outcomes in terms of meeting effectiveness, comfort, and inclusiveness. This is supported by qualitative feedback that mostly favours the head and face motion of webcam-animated avatars over voice-animated lip-syncing only or a static picture. Qualitative results are presented as a systematic thematic framework that allows for comparison of the effects of avatar animation modality preferences and meeting outcomes. We discuss the implications of these results for research and commercial deployment, arguing that webcam-animated avatars are a plausible alternative to video in work meetings, albeit with some nuances, and that visual fidelity is less important than motion fidelity for meeting outcomes. In summary, this paper's key contributions are:
\begin{itemize} 
\item An empirical comparison of avatar animation modalities in terms of their impact on meeting outcomes and user preferences in a controlled study with ecological validity (participants were employees of a global technology company who had worked with one another, and tasks were comparable to work scenarios). This study, designed to replicate the intricacies of professional interactions, presented a unique challenge in recruitment due to the involvement of individuals from diverse corporate backgrounds. 
\item Quantitative and qualitative evidence for webcam-animated avatars providing the highest support for decision agreement in work meetings, suggesting that they are a plausible alternative to video in work meetings. 
\item A systematic thematic qualitative framework of ten factors of user responses about avatars in three categories, which allows for comparison of factors around user preferences versus those around meeting outcomes. 
\end{itemize}

\section{Background and Related Work}
\label{SEC:Background}

While video in online meetings is valued for its apparent richness of expression, its use is not without problems. Situational issues with using video include desiring privacy, confidentiality or anonymity, coping with low bandwidth, or an inability to use video due to a need for mobility \cite{karl_goodbaduglyvideo_2022, castelli_studentvideo_2021, brubaker_beyondcamera_2012}. Other issues are more personal, such as appearance anxiety \cite{kuhn_constant_2022, castelli_studentvideo_2021,Laboratory} or physical, cognitive, emotional burden due to disability or neurodiversity \cite{tang_understanding_2021, das_towards_2021}. Many of these issues also contribute to videoconferencing fatigue \cite{riedl_stress_2022, bailenson2021nonverbal}. Video-on meetings have become a source of tension between those who wish to use it to establish inclusion \cite{cutler_inclusion_2021}, which involves rapport, presence, or attention, and those who cannot or will not use video \cite{bergmann_meeting_2023, kuzminykh_lowattentiondeliberate_2020, weller_videoininterviews_2017}. One solution to expand beyond the binary choice of turning video on or off in online meetings is to use avatars for visual representation of participants \cite{panda_avatarmeetingalltogether_2022}. \citet{ratan_facial_2022} suggest that avatars may reduce videoconferencing fatigue by occluding the user's video and thus reducing self-focused attention. However, the question remains as to how this will affect meetings themselves. In the following sub-sections, we review prior work which has investigated the appearance, realism, and animation of avatars, as well as their use in professional and collaborative settings (\autoref{subsec:appearance}). We also review relevant research focused on measuring aspects of avatar satisfaction and meeting outcomes (\autoref{sec:LRfactor}). Based on this, we propose our research questions and hypotheses (\autoref{RQs}).

\subsection{Avatars: Appearance, Animation, Professionalism, and Use in Collaboration}
\label{subsec:appearance}
 
\paragraph{Appearance and realism} Research on avatars for meetings stretches at least from the late 1990s \cite{davis_avatars_2009}, picks up in the 2010s \cite{shami_avatarmeeting_2010, inkpen_avatarmeeting_2011, jalil_avatarmeeting_2012, Junuzovic2012ToSee}, and retains a strong focus on evaluating the relationship between realism and presence \cite{weidner_realism-hmd-slr_2023, latoschik_avatarrealism_2017, davis_avatars_2009, lombard_at_1997}. Research has found that increased avatar realism does not always correspond to heightened communicative efficacy \cite{nowak_avatars_2018}, and that, in fact, there is a threshold in avatar realism beyond which the enhancement of social presence plateaus \cite{nowakruah_influence_2005}. On the other hand, personalized avatars can foster a deep sense of self-identification and self-presence \cite{YeeBailensonDucheneaut2009, VasalouJoinsonBänzigerGoldiePitt2007}. 

Since realism is very hard to accomplish, customised stylized avatars, which blend abstraction with representation, are easier to create and deploy, but this depends on whether such avatars are regarded as appropriate for work scenarios. In the 2010s, workers were not ready for stylized avatars for work meetings \cite{Junuzovic2012ToSee, inkpen_avatarmeeting_2011} and strongly preferred video or at least more realistic avatars. More recently, the trend appears to be shifting, although not without complications. Anecdotally, during the COVID-19 pandemic, individual experiments with add-ons such as virtual webcams and filters \cite{snapchat_avatars_2023} that produce the effect of avatars for a specific user in a video conference had infamous unprofessional complications \cite{wakefield_notacat_2021, barr_potatoboss_2020}. However, in 2022, \citet{dobre_2022} found that avatars do not necessarily require advanced realism for work meetings (in Mixed Reality). Users can become accustomed to cartoon avatars over a few days, and may even prefer them. \citet{phadnis2023avatars} report a 2023 survey of 2,509 knowledge workers responding that while less realistic avatars may still have acceptance issues, modern realistic avatars also suffer from uncanny valley effects. They argue that the optimal avatar for knowledge work would be stylized with a professional appearance. 

\paragraph{Animation} Beyond stylization and customization, other considerations include the ways in which an avatar moves and the signals which it uses. Although the cutting edge research into producing visual fidelity is exciting~\cite{Li_NERFavatars_2023}, outcome-oriented research on avatars has consistently shown that motion fidelity is more important than visual fidelity \cite{wang_uncanneyvalley_motion_2015}. This holds for both self-identification~\cite{mar_enfacement_motion_2020} and trust in others in meetings~\cite{panda_avatarmeetingalltogether_2022}. Similarly, and in line with the current study's focus on meeting outcomes, decades of videoconferencing research has shown that while people \textit{like} video, good audio is the most important \textit{functional} communicative requirement \cite{isaacs_what_1994, standaert_how_2021}. Video can be heavily perturbed, frozen, or non-existent and a video call will still be possible \cite{watson_good_2000, rintel_conversational_2010}. As such, it should be possible for a conversation to occur between meeting participants with just a picture of a customized avatar as long as the audio is operating. This is very similar to the way that most commercial videoconferencing systems display a picture of audio-only participants in videoconferencing. Audio signals can quite easily drive animation in the form of mouth movement. This may range from simple lip-flapping \cite{bailenson_gaze_lipflapping_2002}, through to viseme shaping of the mouth \cite{tang_visemes_2008}, and even other facial features \cite{li_review-audio-avatar-animation_2022,saunders2023read}. \citet{dobre_2022} used simple lip-flapping along with canned eye-blinking to achieve minimally naturalistic motion in their study of cartoon versus realistic avatars in workplace meetings.

Non-verbal movement is key to rich communication and core to our identity. Beyond the obvious component of facial expressions for displaying emotion \cite{burgoon2021nonverbal}, head motion turns out to be the next most important movement component, crucial for persuasion \cite{wells_head_persuasion_1980}, understanding of talk content \cite{munhall_head_understanding_2004}, and both personal and cultural identity \cite{harrison_head_functionculture_2014}. While previous research has explored low-compute methods like IMUs on head-worn devices (like headphones~\cite{Panda2023BeyondAudio}) to drive avatar head movement, using webcam-driven animation allows the use of facial anchors and blendshapes to drive mouth movement and facial expression \cite{wood_3d_2022}. Additionally, body pose for head motion is a well-solved problem \cite{zheng_bodyposereview_2023}. While more complex methods can animate near photorealistic avatars \cite{Li_NERFavatars_2023}, these are still nascent and computationally costly, so simple animation is most scalable for commercial systems.

\paragraph{Professional settings} Previous studies on the use of stylized avatars in professional situations are limited, as are studies conducted with actual organizational employees (as opposed to students). Historically, stylized avatars have been considered inappropriate for work meetings. Users have reported being worried about avatars looking ``unprofessional'' (e.g \cite{Junuzovic2012ToSee}). However, trends may be changing. For instance, a 2022 study by \citet{BusinessMeetings} used a virtual meeting platform that offers avatars representing professional roles like secretaries and officers, exploring one-on-one managerial meetings with avatars. \citet{abdullah_videoconference_2021_[b]} explored the potential for personalized avatars in enhancing user engagement and facilitating more meaningful interactions in professional contexts. In that study, factors such as user preferences, psychological impact, and the establishment of a virtual identity that aligns with an individual's professional persona were key to the effectiveness of avatars. As reported above, \citet{dobre_2022} found that both cartoon and realistic avatars could be used to hold real work meetings in Mixed Reality, although they did not use a specific measure of success. ~\citet{phadnis2023avatars} have surveyed 2,509 knowledge workers about avatar acceptance. They report that while less realistic avatars may still have acceptance issues, hyper-realistic avatars also suffer from uncanny valley effects, and thus argue that the optimal avatar for work contexts is an illustration with professional face, hair, and clothing styles.

\paragraph{Collaboration} In terms of collaboration, avatars with non-verbal cues such as facial expressions and gestures have been found to enhance brainstorming and negotiation tasks among younger users, suggesting that the interactivity and expressiveness of avatars can be beneficial in certain collaborative contexts \cite{ang_effects_2013_[a]}. Similarly, comparing communication patterns in videoconferences and embodied VR environments, research indicated that the medium of interaction influences the dynamics of communication. In VR settings, avatars can offer a more embodied and engaging experience, potentially leading to different outcomes in teamwork and collaboration \cite{abdullah_videoconference_2021_[b]}. 

\subsection{Avatar Satisfaction and Meeting Outcome Factors} 
\label{sec:LRfactor}

The final point above brings us to what we know about measuring the impact of avatar modalities in videoconferencing. While there has been significant prior research into meeting effectiveness, there is a lack of consensus on its precise measurement \cite{hosseinkashi2023meeting}. Further, many identified factors, such as meeting type, meeting hygiene, or specific meeting features, are not directly applicable to avatar animation modalities \cite{allen_featuresofmeetings_,standaert_how_2021}. Collaboration outcomes using avatars have been reported in previous research (e.g., \cite{ang_effects_2013_[a]}), but, again, there are no standardised metrics. As such, for the purposes of this study, we propose a set of factors drawn from meeting science, videoconferencing research, and avatar research, categorised into two sets: \textit{meeting outcomes} and \textit{avatar satisfaction}. 

\paragraph{Meeting outcome factors.} The two core considerations of meeting outcomes in all meeting types and communication modalities are decision agreement and alignment \cite{nixon_impact_1992,Toupin2017SDM,reinig2003toward}. For simplicity of expression in this study, we term these \textit{effectiveness} and \textit{alignment}. Effectiveness is determined by questions about whether a group makes a final decision and whether any modalities make it easier or harder to reach agreement. Alignment is determined by questions about whether participants align with the group decision. 

With respect to avatars and meeting outcomes, it has been found in research on both avatar self-presentation (e.g.~\cite{freeman_avatarperception_2020}) and videoconferencing fatigue (e.g.~\cite{riedl_stress_2022,Fauville2021Tired}) that a comfortable experience is key to engaging with others. For our study, we consider \textit{comfort} to encompass ease of \textit{expressing} oneself using a particular animation modality, and how \textit{tiring} a given animation modality is. 

The final factor pertaining to meeting outcomes is \textit{inclusivity}. Inclusion in the meeting context \cite{cutler_inclusion_2021,Bluedorn1999Inclu,hosseinkashi2023meeting} is the feeling that oneself and others can contribute to discussion, and it has been found to be a key issue in users' choice to turn video on or off when videoconferencing. Given that avatars may replace video, we propose inclusivity as having two aspects, \textit{taking part} and \textit{considering contributions} (each paired for self and others). 

\paragraph{Avatar satisfaction factors.} There are three key factors integral to avatar satisfaction, which encompass both satisfaction with one's own avatar and the avatars of others. \textit{Self-expressive perception} \cite{yarmand_it_2021_[1]} emphasizes the importance of how users express themselves through avatars. \textit{Other-expression perception} \cite{ang_effects_2013_[a]}, focuses on the perceived expressiveness of others in avatar-mediated settings. Finally, overall avatar \textit{preference} \cite{yang_affective_2023_[3]} pertains to the users' preferences for certain types of avatars, underlining the impact of avatar choice and design on user satisfaction.

In sum, an empirical examination of the impact of avatar-based communication in meetings needs to focus on \textit{meeting outcome} factors (effectiveness, alignment, comfort, and inclusivity) and \textit{avatar satisfaction} factors (self-expressive perception, other-expression perception, and avatar preference). 
The specific items that we used to explore these two categories of factors are detailed in \autoref{tab:factors}.

\subsection{Research Questions and Hypotheses} \label{RQs}

As reviewed above, prior work has shown that representational motion fidelity is important in social contexts, such as for generating trust in others in meetings~\cite{panda_avatarmeetingalltogether_2022}, displaying emotion \cite{burgoon2021nonverbal}, persuasion \cite{wells_head_persuasion_1980}, understanding of talk content \cite{munhall_head_understanding_2004}, and both personal and cultural identity \cite{harrison_head_functionculture_2014}. However, how avatar motion fidelity impacts work meeting outcomes remains unclear. Therefore, our first research question and hypotheses focus on \textit{meeting outcomes}:

\begin{itemize}
    \item [\textbf{RQ1.}] How does avatar animation modality impact meeting outcomes?
\end{itemize}

\begin{description}
    \item[H1a:] Meetings conducted with the \textit{webcam-animated (W)} modality would outperform those with the other two modalities, \textit{static picture (S)} and \textit{audio-animated (A)}, on meeting outcome factors.
    \item[H1b:] Meetings conducted with the \textit{audio-animated (A)} modality would outperform those with the \textit{static picture (S)} modality on meeting outcome factors.
\end{description}

Alongside impact on meeting outcomes, user satisfaction with avatars remains an important and evolving factor for adoption. Prior work has shown that body and head tracking along with facial features increases self-identification with avatars (i.e., the enfacement illusion)~\cite{mar_enfacement_motion_2020}, and in another recent study, avatars with head motion were preferred over those without~\cite{panda_avatarmeetingalltogether_2022}.
More broadly, as society's perception of avatars evolves (e.g., see the appropriateness of stylized avatars in professional settings: 2012~\cite{Junuzovic2012ToSee} vs. 2023~\cite{Laboratory}), it is important to maintain an understanding of user preferences as newer capabilities are introduced, including webcam-driven facial features. We thus propose our second research question and hypotheses, focused on \textit{avatar satisfaction}:

\begin{itemize}
    \item [\textbf{RQ2.}] Which avatar animation modality do people prefer?   
\end{itemize}
\begin{description}
    \item[H2a:] The \textit{webcam-animated (W)} modality would be the preferred modality, compared to the other two modalities, static picture (S) and audio-animated (A).
    \item[H2b:] The \textit{audio-animated (A)} modality would be the preferred modality over the static picture (S) modality.
\end{description}

\section{Methods}
\label{SEC:Methods}

\subsection{Participants and Allocation for Meeting Groups}
We recruited 68 participants from a global technology company to take part in one-hour experimental sessions \footnote{This study was approved by the IRB (Ethical ERP ID 10489, RCT ID 5711) of the global technology company and all participants completed an informed consent procedure prior to participation.}.
Participants were recruited via bulk emails to employees who were located in different offices across various countries (e.g., United States, Spain, Italy, United Kingdom etc.).
Participants who expressed interest were asked to recruit an additional 3-5 employees with whom they had professional familiarity. 
We chose to leverage existing relationships among participants, a choice which intersected with that to create personalized avatars instead of random, generic, or self-created avatars. Since familiarity among participants has the potential to enhance group dynamics \cite{cardon_recorded_2023[f], funder_friendsagreement_1988}, we wanted people to be unique and recognizable among group members. This would reduce uncertainty about who was talking, and reduce the confounding influences of avatar animation with the novelty of both appearance and acquaintance \cite{rintel_maximizing_2007} (e.g. if someone chose an avatar or avatar attributes that were novel to other group members, the group might fixate on those differences as an issue rather than the animation level or the task). We subsequently formed 16 groups, 4 groups with 4 members each and the rest with 5 members. Participant ages ranged from 18 to 64, comprising 44 males and 24 females (see details in supplementary material).

\subsection{Personalized Avatars}
For avatar development, participants submitted one self video and two self photographs. Avatar generation was based on several advanced techniques such as 3D face reconstruction \cite{wood_3d_2022}. The procedure is outlined in \autoref{fig:procedure} (a). The generated personalized avatar had head bones to enable head movements and utilizes blendshapes, a commonly used method for facial animation and tracking. The researchers chose accessories for participants based on their photographs, to provide cues to similarity beyond the facial landmarks. Participants were sent three versions of their avatar in advance, allowing them to experiment with movement and select the avatar that they found most comfortable. This approach aimed to mitigate potential dissatisfaction of avatars, which might arise due to limitations in the generation process (such as different angles of the provided photographs) and to reduce the novelty effects of seeing their avatars. Photographs and videos were used exclusively for creating avatars and deleted immediately after the participant approved an avatar.

\subsection{Animation Implementation for Face and Head Movement}
\begin{figure}[h]
  \centering
  \includegraphics[width=\linewidth]{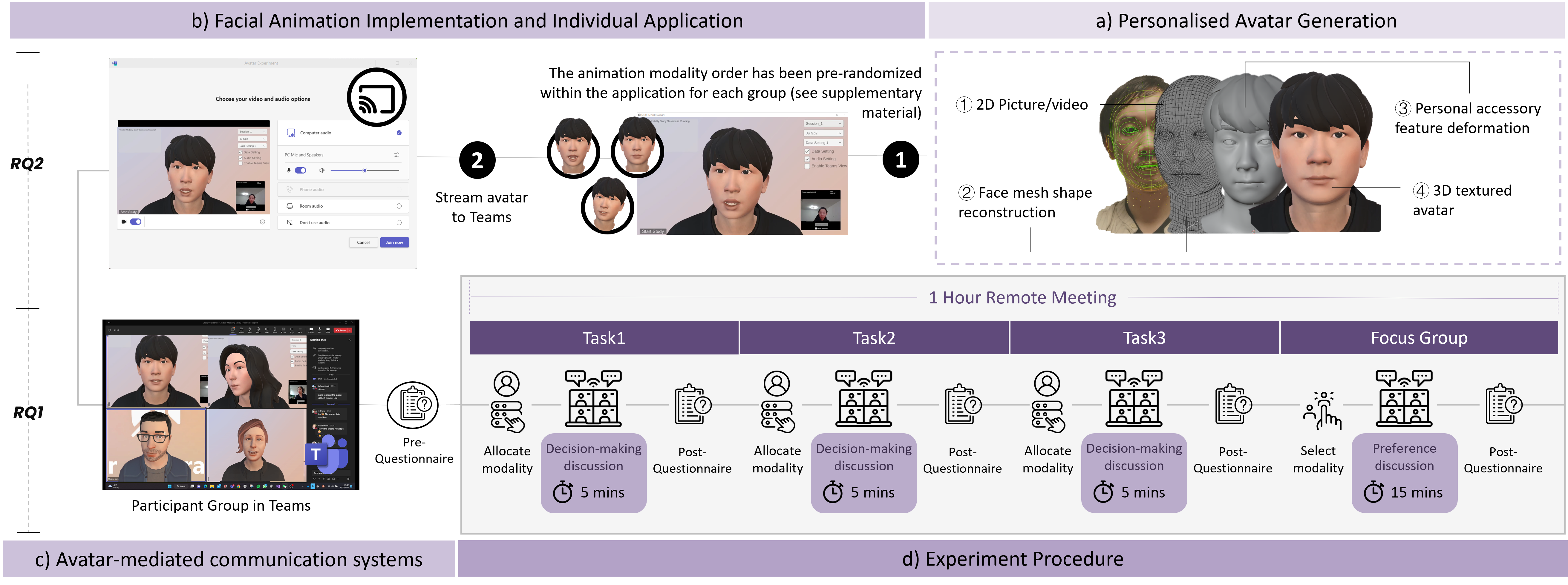}
  \caption{Overview of the experimental protocol based on the research questions. To adequately address RQ2 (avatar satisfaction), (a) personalized avatars were generated based on participants' self videos and photographs. (b) The animation modality order for the avatars was randomized in advance for each group. Avatars were streamed to a Microsoft Teams videoconferencing meeting for the experiment. To address RQ1 (meeting outcomes), (c) participants in a group joined a Microsoft Teams meeting using their avatars to complete the experimental task sessions. Before starting the tasks, participants individually completed an onboarding questionnaire. (d) They then performed three, five-minute group decision-making tasks, each using a different avatar animation modality: static picture mode (S), audio-animated mode (A), and webcam-animated mode (W). Each task was followed by a post-task questionnaire. Lastly, participants completed a 15-minute focus group, for which participants could use any of the three avatar modalities they individually preferred. }
  \label{fig:procedure}
\Description{Overview of the experimental protocol}
\end{figure}

Participants were sent a Unity3D application which enabled them to use their avatar in three animation conditions, and also to change their background to a neutral and clutter-free gradient to avoid attentional confounds. Through Unity Capture\footnote{https://github.com/schellingb/UnityCapture}, the visuals from the application were used as a virtual camera in Microsoft Teams meetings (\autoref{fig:procedure} (b)). In the picture condition, the application just showed a single frame of the avatar in neutral mode for the entire time. In the audio-animated modality, the application used voice for viseme-driven lip animation (blendshape animations). In the webcam-animated modality, the application used video to drive facial expressions (blendshape animations) and head motion (head pan-tilt-yaw).

\subsection{Procedures}

The experiment was conducted using the following procedures, outlined in \autoref{fig:procedure} (d). Participants were asked to complete an onboarding questionnaire at the start of their experiment. The experiment began with three decision-making tasks, each using a distinct avatar animation modality (static picture, audio-animated and webcam-animated). We used the Latin square \cite{latinsquare} design to counterbalance the order of avatar modalities used in the tasks systematically across groups (i.e., groups differed in the order of avatar modalities used in the tasks; see order layout in Supplementary materials), minimizing the influence of presentation order on their responses for each task. The order of the tasks was fixed, as tasks were carefully structured to progress from relatively simple to more selective. This design was intended to eliminate fatigue effects, as in this setup, if the fatigue effect was significant, participants might be more inclined to consider audio during the final focus group.

All task sessions were presented in a gallery format for consistency and a male experimenter acted as a moderator throughout, using his own personalized avatar. All three sessions started with a warm-up question. The warm-up questions were thematically related to the subsequent group decision-making tasks. The warm-up reduced some of the immediate novelty of each animation condition, and also helped familiarise participants with the topic. As highlighted by \citet{groupalignment2007}, group alignment effects can also prime participants for specific types of thinking, leading to more effective and focused group discussions and outcomes in the main tasks. 

Warm-up questions were followed by a 5-minute group decision-making task. The group decision-making tasks had three different emphases from McGrath’s circumplex \cite{abdullah_videoconference_2021_[b]}, which models different forms of group interaction. The emphases were 
Creativity, Planning, and Decision Rationale. The difference in emphases provided a variety of engagement with the decision-making component to reduce the potential for boredom or repetition for the participant groups. The decision-making tasks were as follows:

$\textit{Q}_{\textit{\small{Task1}}}$: If you were going to put one thing in your organization’s welcome pack for new starters, what would it be? Assume a budget of \$50 per person. (Creativity emphasis) 

$\textit{Q}_{\textit{\small{Task2}}}$: Imagine you’re planning one social gathering for your team, what type of activity would that be? Assume a budget of \$100 per person. Ignore the geography constraint. Please consider accessibility needs of people in the organisation. (Planning emphasis) 

$\textit{Q}_{\textit{\small{Task3}}}$: Imagine you are setting up a new <organization> office, where would that be and what is the most important factor in making this decision? (Decision Rationale emphasis) 

After each task session, participants completed a 3-minute survey about \textit{meeting outcome factors} ($\textit{Q}_{\textit{\small{MO}}}$) and \textit{avatar satisfaction factors} ($\textit{Q}_{\textit{\small{AS}}}$), as outlined in ~\autoref{tab:factors}. 
The decision-making task sessions were followed by a 15-minute focus group session. Instead of being allocated an avatar animation modality in the focus group, users were empowered to choose their preferred modality from any of the three modalities they had just experienced, which provided an additional preference metric beyond those captured in the task sessions. The focus group was tailored to extract qualitative insights into participants' experiences, preferences, and feedback. Finally, as the focus group ended, participants were offered the opportunity to complete an \textit{optional} post-session questionnaire, which served as supplementary data for participants to voice any thoughts or concerns they may have been reluctant to share during the group discussion. The entire experiment lasted approximately one hour. 

\subsection{Quantitative and Qualitative Methods} 

\subsubsection{Quantitative Methods} 

As noted in the review of prior work on avatars and meeting outcomes (\autoref{sec:LRfactor}), our measurement of \textit{meeting outcome} factors included the assessment of \textit{effectiveness}, \textit{alignment}, \textit{comfort}, and \textit{inclusiveness}. The majority of these questions were adapted from prior work. Questions on comfort and inclusiveness were adapted from \citet{Cutler2021ComfortInclu} and \citet{Bluedorn1999Inclu}. Comfort ($\textit{Q}_{\textit{\small{MO-comf}}}$) was measured with a 5-point response scale ranging from `Very uncomfortable' to `Very comfortable'. The question on meeting fatigue was adapted from \citet{Fauville2021Tired}. Inclusiveness ($\textit{Q}_{\textit{\small{MO-incl}}}$) consisted of four questions measured with a 5-point response scale (ranging from `Not at all' to `Very much').  
We created questions for effectiveness and alignment that suited the decision-making tasks (partly inspired by \cite{Toupin2017SDM,reinig2003toward}). Effectiveness ($\textit{Q}_{\textit{\small{MO-eff}}}$) was presented as a binary Yes/No question. Alignment ($\textit{Q}_{\textit{\small{MO-align}}}$) was measured using a 5-point Likert scale.
Avatar satisfaction ($\textit{Q}_{\textit{\small{AS}}}$) was measured by asking participants about the perceived effectiveness of the avatars in terms of their self-expression ($\textit{Q}_{\textit{\small{AS-self}}}$) and the expression of other team members ($\textit{Q}{\textit{\small{AS-other}}}$), each measured on a 5-point response scale ranging from `Not at all' to `Very much'. Finally, participants' \textit{preference} ($\textit{Q}_{\textit{\small{AS-pref}}}$) was measured by their actual choice of avatar animation modality for the focus group.  These questions were integrated into task questionnaires. Specific questions can be found in ~\autoref{tab:factors}.

\begin{table}[ht]
\centering
\begin{tabularx}{\linewidth}{p{2cm} p{1.7cm} p{1.9cm} X@{}}
\toprule
\multicolumn{4}{c}{\textbf{Meeting Outcome Factor} ($Q_{\textit{MO}}$)} \\
\midrule
\midrule
\textbf{$Q_{\textit{MO}}$} & \textbf{Method} & \textbf{Measure} & \textbf{Question} \\
\midrule
\multirow{2}{*}{\textbf{\textit{effectiveness}}} 
  & Quantitative & Yes/No & $Q_{\textit{MO-eff}}$ : Did the group make a final decision? \\
\cdashline{2-4}
  & Qualitative & Free text & $Q_{\textit{MO-eff-text}}$ : Did any modality make it easier or harder to reach agreement? Why? \\
\cdashline{1-4}
\textbf{\textit{alignment}}
  & Quantitative & 5-point response scale & $Q_{\textit{MO-align}}$ : After the discussion, how much do you agree or disagree with the group’s decision? \\
\cdashline{1-4}
\multirow{2}{*}{\textbf{\textit{comfort}}}
  & Quantitative & 5-point response scale & $Q_{\textit{MO-comf}}$ : How comfortable did you feel expressing yourself via this visual modality? \\
\cdashline{2-4}
  & Qualitative & Free text & $Q_{\textit{MO-comf-text}}$ : Were any modalities more or less tiring than others? \\
\cdashline{1-4}
\textbf{\textit{inclusivity}}
  & Quantitative & 5-point response scale &
\begin{tabular}[t]{@{} X @{}}
  $Q_{\textit{MO-incl}}$ : \\
  1. To what extent do you feel you had a chance to take part in the discussion? \\
  2. To what extent do you feel others had a chance to take part in the discussion? \\
  3. To what extent do you feel you considered contributions from others? \\
  4. To what extent do you feel others considered your contributions?
\end{tabular}\\
\midrule
\multicolumn{4}{c}{\textbf{Avatar Satisfaction Factor} ($Q_{\textit{AS}}$)} \\
\midrule
\midrule
\textbf{$Q_{\textit{AS}}$} & \textbf{Method} & \textbf{Measure} & \textbf{Question} \\
\midrule
\textbf{\textit{self-expressive perception}}
  & Quantitative & 5-point response scale & $Q_{\textit{AS-self}}$ : To what extent do you feel your representation was effective in expressing yourself? \\
\cdashline{1-4}
\textbf{\textit{other-expression perception}}
  & Quantitative & 5-point response scale & $Q_{\textit{AS-other}}$ : To what extent do you feel the representation of others was effective in expressing themselves? \\
\cdashline{1-4}
\multirow{2}{*}{\textbf{\textit{preference}}}
  & Quantitative & Choice from 3 modalities & $Q_{\textit{AS-pref}}$ : The actual avatar animation modality choice participants made to represent them in the focus group. \\
\cdashline{2-4}
  & Qualitative & Free text & $Q_{\textit{AS-pref-text}}$ : Why did you choose to appear in your current modality? \\
\bottomrule
\end{tabularx}
\caption{Meeting outcome and avatar satisfaction factors measurement and questions. All quantitative questions were asked in a survey after each task conducted in the different modality. All qualitative questions were asked in the final optional post-questionnaire after the Focus Group session. For \textit{comfort}, the response scale ranged from `Very uncomfortable' to `Very comfortable'. For \textit{inclusivity}, \textit{self-expressive perception}, and \textit{other-expression perception}, it ranged from `Not at all' to `Very much'.}
\label{tab:factors}
\end{table}

\subsubsection{Qualitative Methods}
\label{method:qual}

Incorporating qualitative findings alongside our quantitative findings enabled a nuanced understanding of participant reasoning for their choices and unpacking distinctions between personal preferences and meeting outcomes.
Qualitative analysis was conducted on the free text questions from the questionnaires (as shown in ~\autoref{tab:factors}). There were three free text questions relating to \textit{preferences} ($\textit{Q}_{\textit{\small{AS-pref-text}}}$), \textit{effectiveness} ($\textit{Q}_{\textit{\small{MO-eff-text}}}$) and \textit{comfort} ($\textit{Q}_{\textit{\small{MO-comfort-text}}}$),  all factors are crucial for $\textit{Q}_{\textit{\small{MO}}}$. 
The questions were presented after participants had chosen their avatar modality for the focus group, providing detailed reasoning for their preference ($\textit{Q}_{\textit{\small{AS-pref-text}}}$) and how the different modalities influenced meeting outcomes ($\textit{Q}_{\textit{\small{MO-eff-text}}}$ and $\textit{Q}_{\textit{\small{MO-comfort-text}}}$).

Initially, we used a bottom-up inductive thematic analysis method \cite{qualitative2013} starting with the \textit{preferences} data ($\textit{Q}_{\textit{\small{AS-pref-text}}}$). 
Themes were derived from the data, starting with low-level codes, which were then combined into higher-level themes, allowing the data to guide the identification of bottom-up patterns. However, when this approach was applied to the $\textit{Q}_{\textit{\small{MO}}}$ questions ($\textit{Q}_{\textit{\small{MO-eff-text}}}$ and $\textit{Q}_{\textit{\small{MO-comf-text}}}$) it was clear that we had reached saturation, in the sense that codes from the new data were a good fit with existing themes rather than introducing new themes. We nonetheless continued with the analysis to understand which of the existing themes were important for \textit{meeting outcomes}. We therefore then took a top-down, deductive approach in which we used the existing themes as pre-defined codes which were applied to the data, rather than extracting new codes from the data. This enabled us to analyse where participants discussed these themes in relation to \textit{effectiveness} and \textit{comfort}.

We used NVivo 12 to assign quotes and generate initial codes from post-questionnaire text (see coding process in supplementary material). We then identified themes from these codes, which helped us distill meaningful insights from the participants' responses. NVivo 12 also facilitated effective code counting, which helped us understand the relative frequency of themes and compare individuals' perceptions of Avatar and Meeting Outcome. Thus, code counts generated during the coding process were used to assign meaning to specific portions of data. Some quotes are counted multiple times because they encompass multiple meanings assigned to different codes. For instance, when a participant reported that \textit{``the [webcam-animated modality] was the most expressive, and made me feel most `like myself' in a meeting,''} it encompassed both the expression of avatars and the sense of resembling one's self in the representation. In this situation, this quote will be counted twice in two different themes. 

\section{Results}
\label{SEC:Results}

We first present our quantitative analysis for meeting outcome and avatar satisfaction factors. We then present a qualitative thematic analysis for the responses provided by textual questions after the focus group sessions. 

\subsection{Quantitative Results}

Our quantitative results (\autoref{tab:result}) cover four \textit{meeting outcome} factors and three \textit{avatar satisfaction} factors, based on surveys conducted immediately after each of the three task sessions. Other than \textit{effectiveness}, which is a Yes/No question, all the other questions were measured by a 5-point  scale. For \textit{alignment}, the response scale was a Likert scale. For \textit{comfort}, the response scale ranged from `Very uncomfortable' to `Very comfortable'. For \textit{inclusiveness}, \textit{self-expressive perception}, and \textit{other-expression perception}, it ranged from `Not at all' to `Very much'. All tests were conducted in SPSS version 27.   

\subsubsection{Avatar Animation Modality and Meeting Outcome Factors ($\textit{Q}_{\textit{\small{MO}}}$)} \label{Result:quan}

For \textit{effectiveness}, after each modality, each participant independently indicated whether they thought the group had reached an agreement (Yes/No answer). Sixty-five participants indicated that they had reached agreement in the webcam-animated modality (96\%), 59 for the audio-animated modality (87\%), and 54 for the picture modality (79\%). A Test of Proportions (using our custom Matlab function, see supplementary material)
revealed a significant difference between webcam-animated \& picture modality ($ z = 2.85, \textbf{p = 0.004}$), but not between webcam \& audio-animated modality ($z = 1.81, p = 0.070$) nor audio \& picture modality ($z = 1.14, p = 0.25$). This suggested that in the webcam-animated modality, more participants thought they reached group agreement compared to the picture modality. Each participant also indicated their level of \textit{alignment}, \textit{comfort}, and \textit{inclusivity} after each modality on a 5-point scale. $\textit{Q}_{\textit{\small{MO-align}}}$ \& $\textit{Q}_{\textit{\small{MO-comf}}}$  were both single-item questions. For $\textit{Q}_{\textit{\small{MO-incl}}}$ , we took the average of the four questions. All four measurements failed the Shapiro-Wilk test for normality ($p<0.05$), so we used the Wilcoxon Signed-Rank Test (non-parametric equivalent to paired t-test). 

\begin{figure}[h]
    \centering
    \includegraphics[width=1\linewidth]{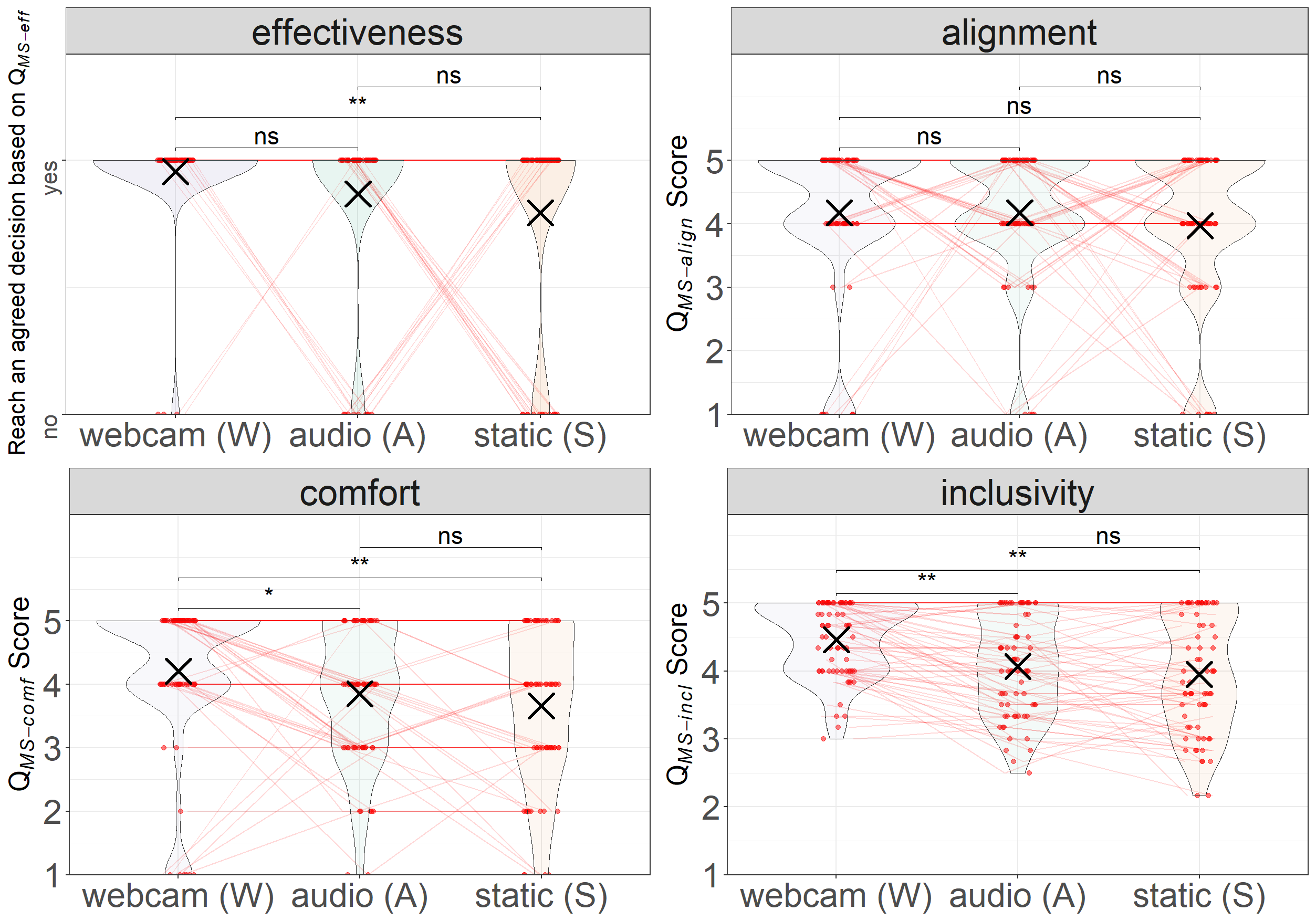}
    \caption{Quantitative results comparing the three avatar modalities on the outcomes of meeting effectiveness, alignment, comfort, and inclusivity. Violin plots depict distributions of numeric data using density curves. Red dots and lines depict individual responses, illustrating comparisons across the three modalities. Black crosses indicate mean values. For effectiveness, a yes/no response scale was used to measure whether the meeting reached an agreed decision. For alignment, comfort, and inclusivity, a five-point response scale was used, where 1 indicates `low' and 5 indicates `high' on the response scale, respectively. * indicates p < 0.05, ** indicates p < 0.01, `ns' indicates not statistically significant.}
    \Description{Quantitative results}
    \label{fig:msresult}
\end{figure}

As shown in ~\autoref{fig:msresult}, for \textit{alignment}, there was no difference between the three conditions. We also ran tests excluding participants who said that their group did not reach an agreement in \textit{effectiveness}, and received a similar result. For \textit{comfort} \& \textit{inclusivity}, the webcam-animated modality was perceived to be the best. However, although the audio-animated modality was higher in both measurements than picture modality, the differences were not significant. 

Overall, our \textit{meeting outcome} factors results supported \textbf{H1a}, in terms of \textit{comfort}, \textit{alignment}, and sense of \textit{inclusivity}, suggesting that the \textit{webcam-animated} (W) modality outperformed the other two modalities in general. However, in terms of \textit{effectiveness}, the \textit{webcam-animated} (W) modality was statistically better than the static picture (S), but only marginally better than audio-animated (A). Also, in terms of \textit{alignment}, no statistical differences were found between any of the modalities. \textbf{H1b} was not supported, as we found no difference between audio-animated and picture modality, in any of the \textit{meeting outcome} factors measurements. 

\hfill\\

\begin{table}
\centering
\begin{tblr}{
  width = \linewidth,
  colspec = {Q[135]Q[83]Q[65]Q[73]Q[8]Q[190]Q[190]Q[181]},
  cell{1}{2} = {c=3}{0.2\linewidth,c},
  cell{1}{6} = {c=3}{0.5\linewidth,c},
  cell{2}{2} = {t},
  cell{2}{3} = {t},
  cell{2}{4} = {t},
  cell{2}{6} = {c,t},
  cell{2}{7} = {c,t},
  cell{2}{8} = {c,t},
  cell{3}{2} = {c=3}{0.2\linewidth,c,b},
  cell{3}{6} = {c=3}{0.5\linewidth,c,b},
  cell{5}{2} = {c=3}{0.2\linewidth,c},
  cell{5}{6} = {c=3}{0.5\linewidth,c},
  cell{4}{7} = {gray!15},
  cell{7}{6} = {gray!15},
  cell{7}{7} = {gray!15},
  cell{8}{6} = {gray!15},
  cell{8}{7} = {gray!15},
  cell{9}{6} = {gray!15},
  cell{9}{7} = {gray!15},
  cell{9}{8} = {gray!15},
  cell{10}{6}= {gray!15},
  cell{10}{7}= {gray!15},
  cell{10}{8}= {gray!15},
  hline{1,11} = {-}{0.08em},
  hline{3} = {2-4,6-8}{},
  hline{3} = {2}{2-4,6-8}{},
  hline{4,6} = {2-4,6-8}{0.03em},
  hline{5,7-8,10} = {-}{dotted},
  hline{9} = {-}{0.05em},
}
                          & \textbf{Avatar Animation Modality}                      &       &         &  & \textbf{Comparison} &                               &                             \\
                          & webcam                                   & audio & static &  & webcam vs audio                                           & webcam vs static             & audio vs static            \\
                          & {Proportion \\of effective meetings    } &       &         &  & Test of Proportions                                       &                               &                             \\
\textbf{\textit{effectiveness}}    & 96\%                                     & 87\%  & 79\%    &  & z= 1.81, p = .070                                         & z = 2.85, p = .004**   & z = 1.14, p = .25             \\
                          & Median                                      &       &         &  & Wilcoxon Test                                             &                               &                             \\
\textbf{\textit{alignment}}        & 4.18                                     & 4.18  & 3.97    &  & T = 249, z = .35, p = .729                                  & T = 372, z = .96, p = .339      & T = 436, z = .96, p = .335     \\
\textbf{\textit{comfort}}          & 4.21                                     & 3.85  & 3.66    &  & T = 146, z = 2.46, p = .014*        & T = 474, z = 3.05, p = .002**  & T = 179, z = 1.77, p = .076   \\
\textbf{\textit{inclusivity}}      & 4.61                                     & 4.41  & 4.46    &  &T = 581, z = 2.68, p = .007**          & T = 541, z = 2.92, p = .003**   & T = 290, z = .68, p = .496    \\
\textbf{\textit{self-expressive perception}}  & 4.15                                     & 3.34  & 2.93    &  & T = 703, z = .4.94, p = .000***                             & T = 896, z = .5.16, p = .000*** & T = 418, z = 3.00, p = .003** \\
\textbf{\textit{other-expression perception}} & 4.16                                     & 3.41  & 2.96    &  & T = 841, z = 5.10, p = .000***                              & T = 885, z = 5.03, p = .000***  & T = 601, z = 3.07, p = .002** 
\end{tblr}
\caption{Quantitative survey analysis results comparing impact of three avatar animation modalities on meeting outcome and avatar satisfaction factors.}
\label{tab:result}
\end{table}

\subsubsection{Avatar Animation Modality and Avatar Satisfaction Factors ($\textit{Q}_{\textit{\small{AS}}}$)}
Similar to \textit{meeting outcome} factors, after each modality, participants responded to questions about their perceived level of expressiveness for their self-representation (\textit{self-expressive perception}) and representation of others (\textit{other-expression perception}). Both measurements failed the Shapiro-Wilk test for normality ($p<0.05$) so we used the non-parametric Wilcoxon Signed-Rank Test
. As shown in ~\autoref{tab:result}, for both \textit{self-expressive perception} and \textit{other-expression perception}, the webcam-animated modality was perceived to be the best
. Audio-animation was also perceived to be significantly better than a static picture.  

After experiencing all three modalities, participants were free to choose one for the focus group, which we took as an expression of their avatar \textit{preference}. Webcam-animation was most popular, selected by 65 participants (95.6\%), indicating a very strong preference for representation that combines visual and auditory elements. Just two participants opted for audio-animation (2.9\%), and only one participant chose the picture modality (1.5\%). Overall, the \textit{webcam-animated} (W) modality was the most preferred. In questionnaires, \textit{webcam-animation} (W) is preferred over the other two modalities, and \textit{audio-animation} (A) is preferred over the \textit{static picture} (S) modality, supporting \textbf{H2a} and \textbf{H2b}. 

To summarise our quantitative results: we found that webcam-animation was significantly preferred to the picture modality on three factors related to $\textit{Q}_{\textit{\small{MO}}}$ factors, except for alignment, and also significantly more comfortable and inclusive than the other two modalities. There was, however, no statistically significant difference between the audio-animated and picture modalities. For $\textit{Q}_{\textit{\small{AS}}}$ , there was a similar preference for webcam-animation on all measurements compared to the other two modalities. Audio-animation was also significantly preferred over the picture modality for both self- and other-expression, which was not the case for any of the meeting outcome factors.  In the following section we use qualitative findings to explain and contextualize these results.

\subsection{Qualitative Thematic Analysis } 
\label{Result:qual}

To understand the key motivations behind the trends uncovered in the quantitative analysis, we performed topic analysis of the free text comments from the text input in the post-focus group questionnaire. 
For the purpose of this analysis, the meeting outcome factors questions on \textit{effectiveness} and \textit{comfort} were explored through a deductive approach, while the avatar satisfaction factor of \textit{preference} was explored inductively, as explained in \autoref{method:qual}. The decision to narrow our focus to these specific factors, omitting \textit{alignment} and \textit{inclusivity}, was deliberate to maintain a more manageable scope for our study. Despite not encompassing all potential factors, the results obtained from the selected aspects provide sufficient explanatory detail. Participation in the final questionnaire was optional, and had a 73.5\% response rate (50 of 68 participants).

\begin{table}[h!] 
  \begin{center}
  \begin{tabular}{p{0.15cm}p{6.7cm}|p{7cm}}
      \toprule

     &  \textbf{Theme}                                         & \textbf{Initial codes}          \\
    \toprule
    
    \textbf{1} & [Avatar] Naturalness                                              & \textit{real},  \textit{natural}, \textit{life-like} \\
  
    \textbf{2} & [Avatar] Movement control                                            & \textit{control by motion}, \textit{movement}\\
    
    \textbf{3} & [Avatar] Aesthetic                                                 & \textit{looks good}, \textit{face less fat}, \textit{head leaning forward}\\
    
    \textbf{4} & [Social Interaction] Expressiveness                               & \textit{expressive}, \textit{head nod/shake}, \textit{smile}\\
  
    \textbf{5} & [Social Interaction] Non-verbal Communication                     & \textit{dynamic}, \textit{interactive}, \textit{talking timing}\\

\textbf{6} & [Social Interaction] Representation, Personalisation and Identity        & \textit{keep identity}, \textit{personal representation}, \textit{mimic actions}\\
    
\textbf{7} & [Psychological Effects] Engagement and Interest                       & \textit{engaging}, \textit{fun}, \textit{interesting}\\
    
\textbf{8} & [Psychological Effects] Emotional Awareness and Reaction              & \textit{convey emotions}, \textit{see reactions}, \textit{ability to understand others}\\

\textbf{9} & [Psychological Effects] Cognitive Load and Attention                  & \textit{multitasking}, \textit{focus on task}, \textit{lower cognitive}\\
    
\textbf{10} & [Psychological Effects] Psychological Safety and Trust               & \textit{trust}, \textit{privacy with avatar face}, \textit{relatable experience}, \textit{self-aware},\textit{self-conscious},\textit{aware of how I was appearing}\\

    \bottomrule
  \end{tabular}
  \end{center}
    \caption{Initial codes derived from the question `Why did you choose to appear in your current modality?' and identified themes which reorganised into three categories after inductive coding}
  \label{tab:codingword}
\end{table} 

\subsubsection{Inductive Thematic Development ($\textit{Q}_{\textit{\small{AS}}}$)}
\label{Result:inductive}

The inductive coding process drew upon 95 coding references. In~\autoref{tab:codingword}, we show how the initial codes from $\textit{Q}_{\textit{\small{AS-pref-text}}}$ were then organized into ten key themes, which were further into three categories (Avatar, Social Interaction and Psychological Effects) to aid interpretation. As well as considering the perspective of \textit{preference}, categorizing these thematic results in this way can also provide a clearer perspective on the technical concerns in designing avatars. 

\begin{enumerate}
    \item \textbf{Avatar Themes:}
    These themes related to the visual and movement aspects of the avatar. The design of the avatar held significant import in shaping user interactions and experiences (32 coding references).
\begin{itemize}
  \item \textbf{1 - Naturalness}: Participants explicitly stated that their choice for the modality was influenced by its lifelike and natural characteristics with 17 coding references, as captured in the statements: "\textit{I chose 
  [webcam-animated modality] because it was the most life-like}" and "\textit{It's [webcam-animated modality] the most natural one because I can see the feedback from my movements and it just seems more natural.}" Additionally, the concept of `realness' emerged as a dimension. It was not just the real appearance of the avatars that mattered but also an overall experience: "\textit{[webcam-animated modality] closest to real video stream while having movement.}"

  \item \textbf{2 - Movement Control:} Another key reason cited by participants for selecting the webcam-animated modality pertained to the avatar's movement. Specifically, participants emphasised the importance of responsiveness in avatar movement. Illustrative comments include: "\textit{I was constantly using [...] webcam-animated modality] to nod or shake my head to make things seem more natural,}" "\textit{Session 2 [webcam-animated modality] - having my avatar move with my movements is critical vs having a static version or just mouth move - that felt distracting,}" and "\textit{I also found my Session 1 avatar [webcam-animated modality] to be really accurate to my face! I couldn't stop staring at it!}".

  \item \textbf{3 - Aesthetics:} 
  The appearance of the avatars was consistent across three modality conditions, but specific observations were made nevertheless. One participant mentioned that the webcam-animated modality made her feel that her "\textit{Face was less fat}," leading her to opt for this particular modality. More negatively, another participant pointed out that "\textit{My head was leaning forward [in the webcam-animated modality]}," which led him to choose the audio-animated modality
   as it offered a more 
  natural head angle. This leaning forward is likely a result of a technical issue with capturing head pose, as opposed to the designed experience, which shows how sensitive users are to errors. 
\end{itemize}

\item \textbf{Social Interaction Themes:}
This factor encompassed themes related to communication and social interactions within the meetings. Themes involve non-verbal communication, social behaviours, and the role of Avatar Satisfaction factors in facilitating or hindering social interaction. This category underscores the critical role that social interactions play in shaping user experiences (37 coding references).
\begin{itemize}
  \item \textbf{ 4 - Expressiveness:} The word 'expressive' came up a lot when people talked about why they chose a particular modality. Participants exhibited a preference for webcam-animated avatars, as indicated by statements such as, "\textit{I like the expressivity}", "\textit{It is more interactive than the others.}", and 
  "\textit{I feel like the [webcam-animated modality] better represents me and enables me to better express myself with some facial expressions and movements}". Also, quotes like "\textit{I thought I could express myself better.}" are indicative of a connection to the perception of self-expression ($\textit{Q}_{\textit{\small{AS-self}}}$). This theme had the most coding references (17). 

  \item \textbf{5 - Non-verbal Communication:} This theme centred on the significance of non-verbal communication for selecting a modality. While the audio-animated modality offered some visual feedback (a moving mouth while speaking), participants drew attention to the comprehensive advantages offered by the webcam-animated modality. Responses such as, "\textit{We know when people are talking in [webcam-animated modality]}", "\textit{The [picture modality] was somewhat exhausting, as it presented difficulties in discerning the appropriate moments to contribute to the discussion,}" and "\textit{In [picture] and [audio-animated modalities] cannot tell whether people are ready to start or finish talking,}" accentuate the value of dynamic non-verbal cues. Non-verbal communication elements like nodding and smiling were said to provide additional value. One participant explicitly emphasized the utility of the webcam-animated modality by stating, "\textit{I appreciated that I was able to convey emotions and reactions using non-verbal communication (head nodding, smiling, etc) without having to rely on a video call.}"

  \item \textbf{6 - Representation, Personalization, and Identity}: Participants valued modalities that allowed for a better representation of their personality and identity. One noted that "\textit{The [webcam-animated modality] felt like it allowed my personality to come across more in the discussion.}". Similar comments included "\textit{The [webcam-animated modality] was the most expressive, and made me feel most "like myself" in a meeting}" and "\textit{it's important to try new experience not losing the identity so I chose the [webcam-animated modality]}".

\end{itemize}
\item \textbf{Psychological Effects Themes:}
The psychological effects of avatars were expressed in 26 coding references, comprising themes such as engagement, emotional connection, cognitive load and the sense of privacy. Although this had the fewest codes among the three categories, its importance should not be overlooked. Indeed, it holds considerable potential for influencing users' attitudes and behaviours \cite{bailenson_2021_Nonverbal}.

\begin{itemize}
  \item \textbf{7 - Engagement and Interest}: This theme highlighted the importance of a new kind of meeting experience through avatars. Users found it exciting, as one participant noted, "\textit{I like the [webcam-animated modality] animations. They remind me that I'm speaking with real people in real time. I also found my 
  avatar to be really accurate to my face! I couldn't stop staring at it!}" This sense of enjoyment contributed to engaging experiences, affecting choice of modality. This was further corroborated by participant remarks like, "\textit{The [webcam-animated modality] is the most engaging and lifelike.}" "\textit{This [webcam-animated modality] was fun!}" and "\textit{The [webcam-animated modality] seemed like the most interesting one because it is dynamic.}".
  
  \item \textbf{8 - Emotional Awareness and Reaction}: The theme was underscored as an essential reason to choose a modality, especially for improving perception and expression of emotions. Participants appreciated the ability to convey emotions, with statements like, "\textit{I appreciated that I was able to convey emotions and reactions using non-verbal communication by [webcam-animated modality]}," and "\textit{That [webcam-animated modality] makes me feel more comfortable and confident while I can see my peers' emotions.}"

  \item \textbf{9 - Cognitive Load and Attention}: Participants preferred modalities that minimised cognitive load and helped them focus on tasks. Participants mentioned, "\textit{The [webcam-animated modality] encouraged me to stay focused on the task and discussion.}" and \textit{I like the expressively and lower cognitive burden of trying to deduce other people's states / make my own clear by voice only.}
  
  \item \textbf{10 - Psychological Safety and Trust}: This theme stood out even though it was only mentioned four times. Participants said things like, "\textit{I chose the [webcam-animated modality] because it feels more natural and trustworthy,}" and "\textit{I appeared as the [webcam-animated modality] because it felt expressive but didn't invade my privacy like a full-on camera would.}"
\end{itemize}

\end{enumerate}

\subsubsection{Deductive Thematic Analysis (across $\textit{Q}_{\textit{\small{AS}}}$ and $\textit{Q}_{\textit{\small{MO}}}$)}

Since the analysis of \textit{effectiveness} and \textit{comfort} did not introduce new themes by inductive process, we took a deductive approach, mapping the 10 pre-existing themes derived from preference $\textit{Q}_{\textit{\small{AS-pref-text}}}$ to the data from effectiveness $\textit{Q}_{\textit{\small{MO-eff-text}}}$ and comfort $\textit{Q}_{\textit{\small{MO-comf-text}}}$ (excepts from quoted content in \autoref{Result:inductive}; more coding references from $\textit{Q}_{\textit{\small{MO}}}$ can be found in the supplementary material). This allowed us to determine which of the themes were seen as relevant for \textit{effectiveness} and \textit{comfort} while also seeing if there were any themes that were seen as important for \textit{modality preference}, but not because of their influence on \textit{effectiveness} or \textit{comfort}.  

\begin{figure}[h]
  \centering
  \includegraphics[width=0.9\linewidth]{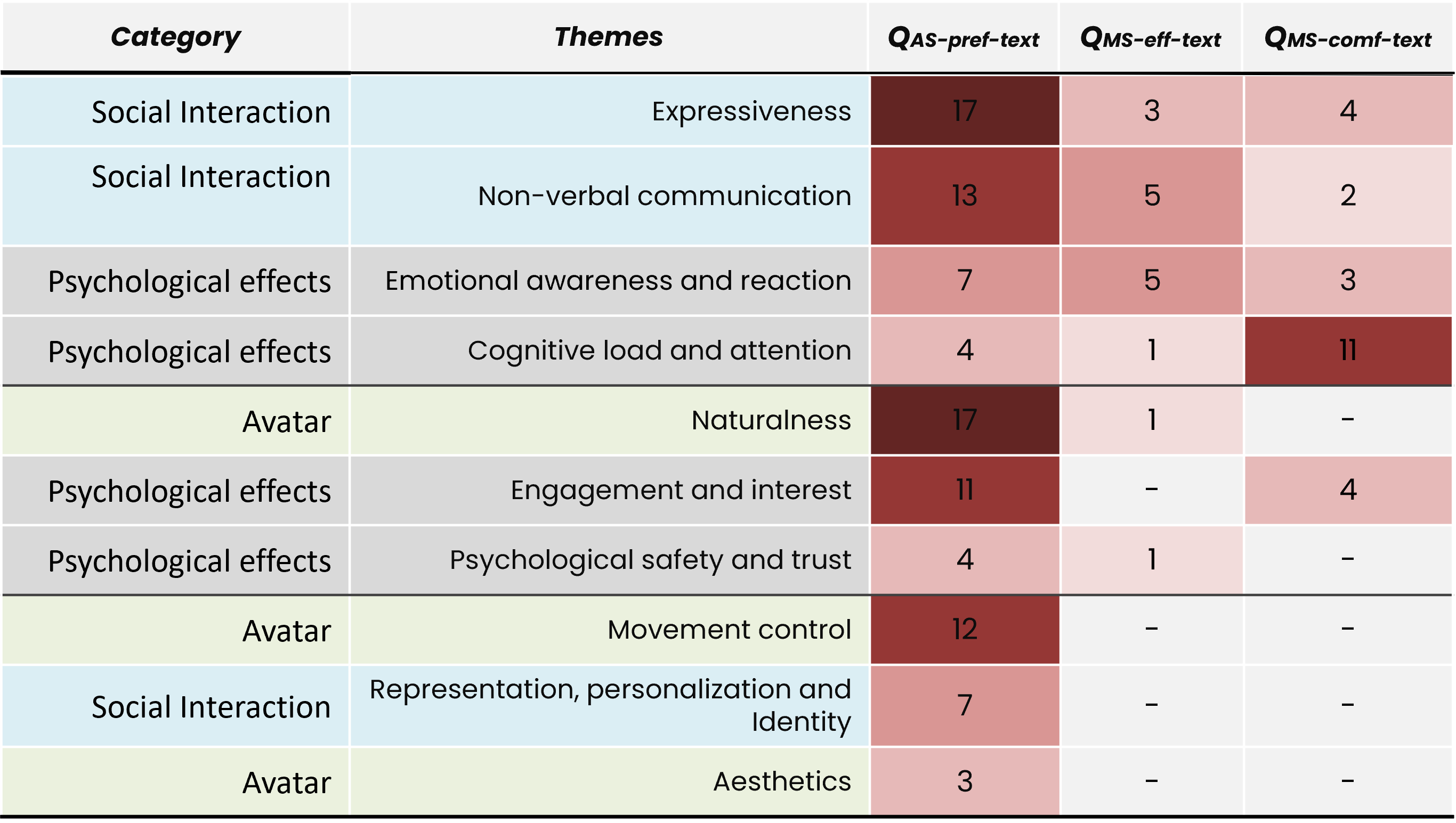}
  \caption{Thematic interrelations between the meeting outcome: \textit{effectiveness} ($\textit{Q}_{\textit{\small{MO-eff-text}}}$) and \textit{comfort} ($\textit{Q}_{\textit{\small{MO-comf-text}}}$) and avatar satisfaction factors: \textit{preference} ($\textit{Q}_{\textit{\small{AS-pref-text}}}$) ranked in descending order in impact themes (primary, secondary and tertiary sections). Additional qualitative Sankey Flow interpretation see supplementary material.} 
  \label{fig:mapping}
  \Description{Thematic interrelations between the meeting outcome}
\end{figure}

\autoref{fig:mapping} shows the results of this analysis. We reordered these themes based on their influence on user preferences and satisfaction which revealed a clear pattern. 
    We identified four \textbf{primary themes} that impacted both Meeting Outcome and Avatar Satisfaction. Primary themes were those that consistently referenced across \textit{all three} three focus group questions: $\textit{Q}_{\textit{\small{AS-pref-text}}}$, $\textit{Q}_{\textit{\small{MO-eff-text}}}$, and $\textit{Q}_{\textit{\small{MO-comf-text}}}$. The four primary themes were \textit{Expressiveness}, \textit{Non-verbal communication}, \textit{Emotional Awareness and reaction}, and \textit{Cognitive Load and attention}.
    We further identified three \textbf{secondary themes}. Secondary themes referenced \textit{any two of the three} focus group questions: $\textit{Q}_{\textit{\small{AS-pref-text}}}$ plus $\textit{Q}_{\textit{\small{MO-eff-text}}}$ or $\textit{Q}_{\textit{\small{MO-comf-text}}}$. The two secondary themes were \textit{Naturalness}, \textit{Psychological safety and trust}, and \textit{Engagement and interest}.
    Three further \textbf{tertiary themes} themes were referenced only in $\textit{Q}_{\textit{\small{AS-pref-text}}}$, calling into question their relevance to Meeting Outcome, as we will discuss later. These themes were \textit{Movement control}, \textit{Representation, personalization, and identity}, and \textit{Aesthetics}.

These findings emphasize the crucial role of communication dynamics on the Social Interaction and Psychological Effects all-avatar meetings. Expressiveness was mentioned 17 times, Non-verbal Communication 13 times, and Emotional Awareness and Reaction 7 times. Although the impact of these themes on the factors of Effectiveness and Comfort was comparatively less pronounced (less than 5 times), people still mentioned it.

Further, the theme of Cognitive Load and Attention, which spans all three satisfaction factors, exhibited an uneven distribution in its mentions. Its prominence in the domain of comfort was most pronounced, with 11 mentions, but much fewer mentions for preference and effectiveness. This disparity suggests that while Cognitive Load and Attention are crucial for meeting \textit{comfort}, their impact on \textit{effectiveness} (1 mention) is relatively limited. Thus, we identify Cognitive Load and Attention and Engagement and Interest emerge as themes influencing both preference and comfort factors, but with less influence on effectiveness. 

Additionally, themes such as Naturalness, Psychological safety, and trust, though mentioned in the context of \textit{effectiveness}, were not as frequently cited. This observation led us to categorize these themes as more specific to avatar preference than to overall Meeting Outcome (similar to the themes of Movement control, Representation, Personalization and Identity, and Aesthetics). These findings indicate that these factors are important for choice of modality but not becuase of their influence on either \textit{effectiveness} or \textit{comfort}. 

\subsubsection{Connections between the Quantitative results and Qualitative analysis}
\label{ref:interpretationQuantQual}

\paragraph{Impact on effectiveness} 
We found that the effectiveness of decision-making varied depending on the type of questions asked. After each task, users were asked $\textit{Q}_{\textit{\small{MO-eff}}}$ ("Did your group make a final decision?") and $\textit{Q}_{\textit{\small{MO-eff-text}}}$ ("Did modality make it easier or harder to reach agreement?"). In our quantitative analysis, a statistically significant number of participants reported that the group agreement was reached with the \textit{webcam-animated} (W) modality compared to the audio-animated (A) and picture (S) modalities. However, in the qualitative analysis, 66\% of participants (33 out of 50) believed that the avatar modality had no influence on their group decision making. To gain a deeper understanding of the dynamics, we analysed the actual quotes.

Participants emphasised support for the impact of avatar animation modality when discussing the role of Naturalness, Expressiveness, Non-verbal communication, Emotional awareness and reaction and Psychological safety and trust in the decision-making meeting contexts. For example, quotes mentioned that "\textit{It [webcam-animated modality] was helpful to see smiles and head nods as feedback to my statements",}  and "\textit{first modality [webcam-animated modality] is for sure the more complete to create a trust with other people joining the call}".

On the other hand, participants de-emphasised the impact of avatar animation modality due to two factors. First, the complexity of the decision-making tasks themselves were said to have a greater influence on the likelihood of reaching an agreement than the modality used. One participant stated, "\textit{I felt the ease of decision-making was more aligned with the question at hand than the modality}." Second, familiarity with one another was said to be important, such that even if they did not have visually expressive cues, they already knew how to collaborate. One participant pointed out, "\textit{Not for this group because we all know each other very well so the non-animated avatars were just like being in a phone call.}"

\paragraph{Comfort and fatigue}
We encountered a range of opinions on the matter of whether a particular modality leads to increased or reduced fatigue. These differing perspectives give rise to distinct expressions in Expressiveness and Cognitive Load and Attention, which can be categorized into pros and cons associated with the tracking that enabled the \textit{webcam-animated} (W) modality. On the pro side (more tracking leads to less fatigue), participants said, "\textit{The [picture modality] session was more tiring because it was only picture therefore requiring a little more work on my end to effectively express myself}" and "\textit{I found the [picture modality] session where folks were frozen to be more tiring. I had to force myself to pay attention to the conversation}" On the con side (less tracking leads to less fatigue), participants said, "\textit{The [webcam-animated] one was the most tiring because it encouraged me to move more.}" and "\textit{Video was more natural for communication and less cognitive load, but I became more self-aware and self-conscious about what my avatar was doing at any given moment (was my mouth open, weird face angle, etc.)}" For the remaining themes (Non-verbal communication, Engagement and interest, the consensus was pro webcam-animation (more tracking lead to less fatigue), with statements such as, "\textit{the pictures were more tiring as we could not see the reactions of the people}". 

\section{Discussion}\label{SEC:Discussion}
In this study, we addressed gaps in the existing literature in understanding the influence of avatar modalities on both meeting outcomes and avatar satisfaction for work videoconferencing meetings, particularly in the context of decision-making.

\subsection{Meeting Outcome Factors}
We hypothesized and found that webcam-animated (W) avatars positively influence meeting outcomes. They have statistically significantly higher scores than the static picture (S) modality for \textit{effectiveness}, and higher than both static (S) and audio-animated (A) modalities for \textit{comfort} and \textit{inclusivity}. Thematic analysis of participants' comments also demonstrated that satisfying one's preferences for avatars is related to expressive qualities best provided with accurate facial and head motion, and these also contribute to meeting effectiveness and meeting comfort. At a base level, then, these findings highlight the potential of webcam-animated (W) animation in enhancing decision-making.

However, the findings about meeting outcomes are more complex. We introduced our interest in meeting outcomes as being whether any modalities make it easier or harder to reach a final decision (\textit{effectiveness}) and easier or harder for participants to align with the group decision (\textit{alignment}). For \textit{effectiveness}, participants' independent indications of whether they thought the group had reached agreement showed significant difference between the webcam-animated (W) and static picture (S) modalities, but not between webcam and audio-animated modalities, nor audio and picture modalities. For alignment, there was no difference between the three conditions. Further, qualitative responses emphasised support for the impact of avatar animation modality for naturalness, expressiveness, non-verbal communication, emotional awareness, and trust-building. However, in the qualitative data, participants also said that the complexity of the decision-making tasks had a greater influence on the likelihood of reaching an agreement than the modality used, and they also said that even when they did not have visually expressive avatars, their pre-existing collegial familiarity meant that they already knew how to collaborate. Thus, whereas our quantitative findings show an impact of avatar modality on some meeting outcomes (i.e., \textit{effectiveness}), participants' perceptions emphasize the relative importance of task complexity and team familiarity for meeting outcomes.

Importantly, our counterbalancing of avatar modalities across tasks eliminates task complexity as a confound in our study. Similarly, our within-subjects experimental design means that team familiarity is fixed across conditions. Thus, our study design accounted for these latter factors, isolating the influence of avatar animation modality on the meeting outcomes of \textit{effectiveness} and \textit{alignment}.  Nevertheless, despite our study's focus on avatar modality, we do not suggest that this factor is \textit{more} important than factors such as task complexity and team familiarity in affecting meeting outcomes. Indeed, participants' comments rightfully acknowledge the complexities of workplace collaboration and social dynamics. Avatar modality may play a relatively smaller yet still influential role in this context. Motivated by participants' comments, future work should explore how team familiarity and task complexity moderate the influence of avatar modality on meeting outcomes.     

Why did participants perceive avatar modality to play a relatively smaller role in meeting outcomes? One possible explanation is that avatar animation modalities affect the social presence and interpersonal trust of the participants, \textit{facilitating} reaching agreement but are not perceived to affect the decision itself. Some older support for this comes from a study by Bente et al.~\cite{bente-2008}, which found that avatars increased the social presence and interpersonal trust of collaborators, especially when they were animated with both video and audio. They suggest that avatars provide nonverbal cues that enhance the impression of being together and the feeling of mutual understanding and reliability, which contribute to decision-making, but are not themselves primary parts of the decision-making process. Our qualitative findings around the value of webcam-animated avatars for social interaction align with this idea. The current study precluded systematically exploring the link between social presence, interpersonal trust, and meeting outcomes due to its limited measurement of meeting \textit{effectiveness} and \textit{alignment}, a result of the study session's time constraints. We leave this important question for future work on avatars.

\subsection{Avatar Satisfaction Factors}
In the context of maintaining consistent personalized avatar appearance, our hypothesis and observations indicate the substantial impact of level of expression on participants' avatar choices for meetings. Notably, there exists a pronounced \textit{preference} for webcam-animated (W) avatars in actual focus group meeting selections (see also \cite{yang_affective_2023_[3]}), with a significant majority (95.6\%). Moreover, our analysis reveals statistically significant differences in both \textit{self-expressive perception} and \textit{other-expression perception} between audio-animated (A) and static picture (S) modality. 
Our results highlight the importance of head movement \cite{wells_head_persuasion_1980, munhall_head_understanding_2004, zheng_bodyposereview_2023} and facial expressions \cite{ang_effects_2013_[a], burgoon2021nonverbal} in the webcam-animated modality, along with mouth motion in both the webcam- and the audio-animated modalities \cite{dobre_2022}. The upshot is that partial or subtle expressive cues can enhance interaction and expressiveness when contrasted with entirely static representations. Additionally, our findings on social interaction align with the literature's emphasis on the importance of non-verbal movement for meetings that benefit from rich communication modalities for optimal outcomes, e.g those to do with building or maintaining relationships, or resolving conflicts~\cite{standaert_how_2021}.

Also in line with prior research, our participants connected their perceived level of self-expression to better self-avatar representation \cite{yarmand_it_2021_[1]} in responses such as \textit{"I feel like it better represents me and enables me to better express myself with some facial expressions and movements."} Within the context of decision-making, perceptions of effectively conveying information to others align with the concept of the expressivity of others~\cite{munhall_head_understanding_2004,ang_effects_2013_[a]}, e.g., in responses such as \textit{"Movements add the ability to guess more body language/nonverbal communication, which is important to understand the feelings of others"} and \textit{"[static picture modality] was a little tiring in the sense that I had a harder time knowing when to hop into the conversation since I couldn't see facial expressions for feedback"}. These observations of avatar satisfaction further imply the pivotal role of modality expressivity in enhancing the overall avatar-mediated communication experience \cite{ang_effects_2013_[a]}.

One tension bubbling somewhat under the surface of the avatar satisfaction results was how participants felt about being seen versus seeing others. Specifically, some people seem to want lower fidelity self-representations, potentially making them `harder to read' (``\textit{The webcam driven one was a bit tiring since I felt like I had to be more `on.'}''), while also wanting higher fidelity representations of others so that they are `easier to read' ("\textit{Movements adds the ability to guess more body language / nonverbal communication, which is important to understand feelings of the others}"). This issue has been reported previously in avatar research: people can feel "exposed" with high-fidelity self-representations, and uncomfortable or uncanny when the movements don't match their own, while preferring high-fidelity representations of others~\cite{panda_avatarmeetingalltogether_2022}. It has also been anecdotally reported in videoconferencing research using standard video, especially during the COVID-19 pandemic, when people reported fatigue about seeing their own video, and thus wanting to turn it off, while also reporting fatigue stemming from not seeing the video of others who have turned their video off~\cite{karl_goodbaduglyvideo_2022}. However, self-representation aversion in videoconferencing is a complex phenomenon. Kuhn~\cite{kuhn_constant_2022} finds that it has correlations with the  dispositional trait of the user's degree of public self-consciousness, while Shockley et al.~\cite{shockley_fatiguing_2021} notes that gender and organizational tenure also correlate with camera usage and fatigue. Future research should conduct a direct test of preferences for self and other representation, and compare avatar modalities to traditional modalities, and check any correlations of modality with both psychological traits and demographic categories.

\subsection{Implications}
\label{SEC:Implications}

Our results have implications for designing avatar-mediated communication systems and remote work meetings.

\subsubsection{Meaningful movement outperforms realism for meeting outcomes}
The \textbf{primary themes} that emerged from the qualitative analysis were all related to motion. Primary themes, as a reminder, are the themes with comments about all three  issues of avatar preferences, meeting effectiveness, and meeting comfort. These themes included Expressiveness, Non-verbal communication, Emotional awareness and reaction, as well as Cognitive load and attention. Collectively, they demonstrate that webcam-animated animation enriches communication by perceived expressiveness and reduces meeting fatigue \cite{allen_featuresofmeetings_, isaacs_what_1994}.
Realism and fidelity are also frequently mentioned in \textbf{secondary themes}, which is in line with previous findings of importance of natural movement \cite{wang_uncanneyvalley_motion_2015,mar_enfacement_motion_2020} for the acceptance of self-expression and expressions of others. That is, the motion of smiles, head nods, and head shakes are considered more critical for achieving effective communication than realism. Our findings suggest that the incorporation of dynamic movements and expressive gestures in webcam-animated modalities can significantly enhance meeting effectiveness by improving comfort and facilitating non-verbal communication. A further study extending our research to delve deeper into the realm of hand gestures would be recommended. Participants themselves expressed this in ways such as, "\textit{Would love to see the incorporation of hand gesture in the future roadmap}" and "\textit{it would be great if the avatar also had hands and could manage reactions with gestures.}"

\subsubsection{Webcam-animated avatars are a plausible alternative in remote work meetings}
As noted in our review of prior work (\autoref{SEC:Background}), historically, stylized avatars have been considered inappropriate for work meetings. The avatars used in our study were stylized, but in our thematic analysis, "Naturalness" stood out prominently, \textit{but only for the webcam-animated avatars}. Indeed, we found a significant difference in user preferences, with webcam-animated (W) avatars being significantly higher rated than audio-animated (A) or picture (S) avatars. The high acceptance of the stylised avatars, providing that they are animated, might be due to the fact that with a rise in the number of platforms that offer avatars, the acceptance of the avatars that are not photo-realistic has also gone up. And, as we note, the most recent survey of avatar acceptance by knowledge workers reported that hyper-realistic avatars suffer from uncanny valley effects. The authors conclude that the optimal avatar for knowledge work should be a stylized illustration \cite{phadnis2023avatars}. Additionally, in our quantitative analysis, the use of webcam-animated avatars resulted in significantly higher \textit{effectiveness}, \textit{comfort} and \textit{inclusiveness} in decision-making. This suggests that in collaborative remote work environments, especially in situations where the utilisation of video is impractical, webcam-animated avatars could be more advantageous compared to merely deactivating the camera.

\subsubsection{Different technical aspects of the experience implementation impact different experience objectives}

Finally, we have found that the top level theme categories that emerged from our thematic analysis -- Avatar characteristics, Social interaction, Psychological effects -- relate to personal user preferences and the success of the meeting in different ways. By understanding how these themes interconnect within these categories, researchers and practitioners can make more informed decisions about crafting avatars that meet the diverse needs and preferences of users. For example, meeting outcomes (\textit{meeting effectiveness} and \textit{meeting comfort}) are mostly satisfied by avatars that support expressiveness, non-verbal communication, emotional awareness and reaction, and cognitive load and attention. At the same time, users' avatar preferences are dominated by themes related to avatar and social interactions (\textit{expressiveness}, \textit{naturalness}, \textit{non-verbal communication}, \textit{movement control}). However, issues of visual realism in terms of representation, personalization, and identity, and aesthetics do not intersect with issues around meeting outcomes. This means that when developing avatars, focusing on visual fidelity might increase the users' immediate satisfaction, but not necessarily lead to a more effective meeting. As we noted in our review of prior work, the functional communicative aspects of meetings rely more on the fidelity of how people move than the fidelity of how they look ~\cite{wang_uncanneyvalley_motion_2015,mar_enfacement_motion_2020,panda_avatarmeetingalltogether_2022}. While developers will need to balance these to ensure that users at least accept their avatars, time with avatars also plays a role. Users can learn to accept less realistic avatars, and even prefer them, as long as they have a motivation to interact using avatars (such as avoiding appearance anxiety, e.g.~\cite{karl_goodbaduglyvideo_2022,kuhn_constant_2022}) and learn that they can accomplish their tasks~\cite{dobre_2022}.

\subsection{Limitations and Future Research}

A key limitation of the study resulted from the trade-off of ecological validity~\cite{kieffer_ecoval_2017} against experimental consistency. We recruited real distributed global teams, but in doing so we had to rely on participants' own technology (including computers, webcams, and internet access) both for conducting the meetings and recording the experimental application. As such, not all groups experienced optimal computer performance, video, and audio quality, and remote debugging of problems was complicated. In order to ensure smooth meeting progress, some data was discarded and some sessions were split into two separate 30-minute slots rather than the ideal single one-hour block.

Given that the experimental setup required coordinating teams, balancing time became a limitation. In order to make the experiment feasible for all participants to attend during their busy work schedule, we had to limit our study (including all three tasks and post-task questionnaires) to a total of one hour, leaving the discussion time for each condition to be capped at five minutes. There were also occasions when one participant had to leave 10 minutes early, or the meeting was inadvertently prolonged. This led to inconsistencies in the duration of some focus group discussions. However, we tried to compensate for these issues in our post-task questionnaire with more textual qualitative data.

Arguably, many of the issue  above are upper-bound limitations. That is, had we been able to control all technology and provide the highest quality conditions, the results may have shown more differentiation between conditions and more significant positive results for animation conditions. However, real-life performance is always likely to be more degraded and subject to both technical and logistical idiosyncrasies. As such, the trade-off of experimental consistency for ecological validity provides results that may enable reasonably robust comparisons in future practitioner and academic work.

A different form of limitation stems from our choice to study teams in which members were already familiar with one another. Some participants noted that this made it easier for them to understand each other during the meeting. Exploring the impact of avatars when participants are unfamiliar with each other (e.g. in situations of onboarding new members to teams or `swift trust' teams) and how this might influence meeting outcomes by using animated avatars would be an intriguing direction to investigate. Combining this with work on existing teams would provide a more comprehensive understanding of the implications of our findings across diverse organizational contexts.

We recognize the potential bias of using data from only one company in our experiment, which may limit the generalizability of our findings. Despite both diversity and cultural dynamics present within this global entity, the restricted sample, unique culture, processes, and organizational structures of this company are nevertheless a constraint. To address this limitation, future research should prioritize expanding the sampled cohort to include data from multiple companies.

We further acknowledge that cultural background and gender could potentially impact the outcome of meetings, and we were unable to recruit a balanced group of participants (24 female, 44 male) from global offices. Out of the 16 groups, three were male-only, and the other 13 were mixed-gender groups (see supplementary materials). However, the impact of gender and culture were not the focus of our current study, and we do not feel that we have enough data to reach any conclusions in these areas. Future work should explore the influence of these areas, for example, given that meeting practices and expectations vary across cultures \cite{kohler_meetings_2015,lehmann-willenbrock_team-meeting_2017}.

Our study focuses on the impact of conditions with all avatars from the same modality. However, in a real-life situation, should participants been given the freedom to choose, it is possible to have mixed avatar modalities in one setting. As we controlled the avatar modality for each discussion conditions, the only mixed-avatar type session appeared in our focus group session, when one of the participants chose anything other than the webcam-animated modality. That is a total of 3 groups (or 3 participants) out of the 16 groups (or 65 participants). Therefore, we do not have enough data to comment on mixed avatar modalities. Future work should be conducted to understand the differences between mixed and single modalities.

Relatedly, our study examined the specific modalities of video-animated, audio-animated, and picture modalities, to provide a baseline understanding of clear differences between animation conditions. However, better overall avatar animation might be achieved by blending animation technologies. There is potential in exploring the mid-level of animation between the video-animated and audio-animated modality. For example, a more natural, expressive audio-animated algorithm could replace simple lip movement replication and add natural head motion. Further, additional non-verbal capture, such as tracking hand gestures tracked, would likely enhance the overall functional communicative value of avatars and have a positive effect on meeting outcomes.

Finally, we introduced the relevance of understanding the impact of avatars in work meetings by noting that avatar usage is nascent in mainstream communication platforms (Zoom and Microsoft Teams have only recently introduced them, while Slack Huddle and Google Meet have yet to introduce them at the time of writing). Avatars driven by similar technologies are further along the adoption journey in entertainment and consumer contexts, such as use by vTubers (video Youtubers) for personal content creation (e.g. exploring performative gender identity~\cite{2021_lu_vtuber-performativeidentity}) and marketing (e.g. avatar marketing  persuasiveness~\cite{2023_sakuma_vtuber-persuasiveness, 2022_debrito_avatarmarketing}). 
These uses are quite different to the professional contexts and outcomes of work meetings, but the technical overlap and salience of identity satisfaction could prompt further exploration into adoption issues and also pave the way for future advancements in virtual collaboration experiences, benefiting both professional and creative domains. 

\section{Conclusion}
\label{SEC:Conclusion}

The findings of this study show support for webcam-animated avatars as an appropriate replacement for webcam video for situations where users do not want to share their live webcam feed. This finding was supported both quantitatively in reports on alignment in meetings using the webcam-animated modality, and in qualitative responses about how specific aspects of webcam animation contributed to aspects of meeting outcomes. Beyond that core finding, our thematic analysis also provided categorizations to consider when tailoring avatar design choices and features to address specific aspects related to avatar appearance, social interaction dynamics, and the psychological impact on users. By understanding how these themes interconnect within these categories, we can make more informed decisions in crafting avatars that meet the diverse needs and preferences of 
users. This categorization serves as a valuable resource not only for researchers seeking a deeper understanding of user preferences but also for industrial designers involved in avatar development.

\begin{acks}
We would like to thank Microsoft for the support of this project, as well as employees from global offices who participated in the experiments. In particular, John Tang and Tadas Baltrusaitis provided valuable experimental suggestions and technical support, while Tom Cashman and Antonio Criminisi offered their support of the overall project. Additionally, Fang Ma, Ju Zhang, Xueni Pan, and Marco Gillies received funding from the Arts and Humanities Research Council (AH/T011416/1).
\end{acks}

\bibliographystyle{ACM-Reference-Format}
\bibliography{sample-base}


\end{document}


\title[Avatar Animation and Meeting Outcome]{Nods of Agreement: Webcam-Driven Avatars Improve Meeting Outcomes and Avatar Satisfaction Over Audio-Driven or Static Avatars in All-Avatar Work Videoconferencing}

\begin{abstract}
Supplementary materials.
\end{abstract}

\author{Fang Ma}
\email{f.ma@gold.ac.uk}
\orcid{0000-0001-5476-6982}
\affiliation{%
  \institution{Goldsmiths, University of London}
  \city{London}
  \country{United Kingdom}
}

\author{Ju Zhang}
\email{j.zhang@gold.ac.uk}
\orcid{0009-0004-3655-4010}
\affiliation{%
  \institution{Goldsmiths, University of London}
  \city{London}
  \country{United Kingdom}
}

\author{Lev Tankelevitch}
\email{lev.tankelevitch@microsoft.com}
\orcid{0000-0003-1286-5194}
\affiliation{%
  \institution{Microsoft Research}
  \city{Cambridge}
  \country{United Kingdom}
}

\author{Payod Panda}
\email{payod.panda@microsoft.com }
\orcid{0000-0002-8208-4185}
\affiliation{%
  \institution{Microsoft Research}
  \city{Cambridge}
  \country{United Kingdom}
}

\author{Torang Asadi}
\email{torang.asadi@microsoft.com}
\orcid{0000-0003-2796-7230}
\affiliation{%
  \institution{Microsoft}
  \city{San Francisco}
  \country{United States}
}

\author{Charlie Hewitt}
\email{chewitt@microsoft.com}
\orcid{0000-0003-3943-6015}
\affiliation{%
  \institution{Microsoft}
  \city{Cambridge}
  \country{United Kingdom}
}

\author{Lohit Petikam}
\email{lohitpetikam@microsoft.com}
\orcid{0000-0001-6629-7490}
\affiliation{%
  \institution{Microsoft}
  \city{Cambridge}
  \country{United Kingdom}
}

\author{James Clemoes}
\email{jaclemoe@microsoft.com}
\orcid{0009-0009-5083-201X}
\affiliation{%
  \institution{Microsoft Research}
  \city{Cambridge}
  \country{United Kingdom}
}

\author{Marco Gillies}
\email{m.gillies@gold.ac.uk}
\orcid{0000-0002-3100-9230}
\affiliation{%
  \institution{Goldsmiths, University of London}
  \city{London}
  \country{United Kingdom}
}

\author{Xueni Pan}
\email{x.pan@gold.ac.uk}
\orcid{0000-0002-1282-1469}
\affiliation{%
  \institution{Goldsmiths, University of London}
  \city{London}
  \country{United Kingdom}
}

\author{Sean Rintel}
\email{serintel@microsoft.com}
\orcid{0000-0003-0840-0546}
\affiliation{%
  \institution{Microsoft Research}
  \city{Cambridge}
  \country{United Kingdom}
}

\author{Marta Wilczkowiak}
\email{mawilczk@microsoft.com}
\orcid{0009-0006-7695-4216}
\affiliation{%
  \institution{Microsoft}
  \city{Cambridge}
  \country{United Kingdom}
}

\renewcommand{\shortauthors}{Ma et al.}

\received{January 2024}
\received[revised]{July 2024}
\received[accepted]{October 2024}

\maketitle

\appendix

\section{Experiment Design}

\subsection{Latin Square counterbalancing method}

\begin{table}[h]
\centering
\resizebox{0.8\textwidth}{!}{%
\begin{tblr}{
  hline{2} = {-}{}}
Order\# & 1st & 2nd & 3rd & Group\# \\
1     & 3 - webcam-animated     & 2 - audio-animated    & 1- static picture    & 1, 7, 13\\
2     & 3 - webcam-animated     & 1- static picture     & 2 - audio-animated    & 2, 8, 14\\
3     & 2 - audio-animated      & 3 - webcam-animated   & 1- static picture    & 3, 9, 15\\
4     & 2 - audio-animated      & 1- static picture     & 3 - webcam-animated    & 4, 10, 16\\
5     & 1- static picture       & 2 - audio-animated    & 3 - webcam-animated    & 5, 11\\
6     & 1- static picture       & 3 - webcam-animated   & 2 - audio-animated      &6, 12 
\end{tblr}
}
\caption{Latin Square layout and distribution of six orders for 16 meeting groups}
\label{tab:latinsquare}
\end{table}
\FloatBarrier

\autoref{tab:latinsquare} represents all six orders of the three animation modality conditions (static picture, audio-animated, webcam-animated), ordered by the balanced Latin Square, ensuring that the conditions are systematically balanced across the meeting groups in the experiment.

\subsection{An interface example of a group meeting discussion}

\begin{figure}[h]
  \centering
  \includegraphics[width=0.65\linewidth]{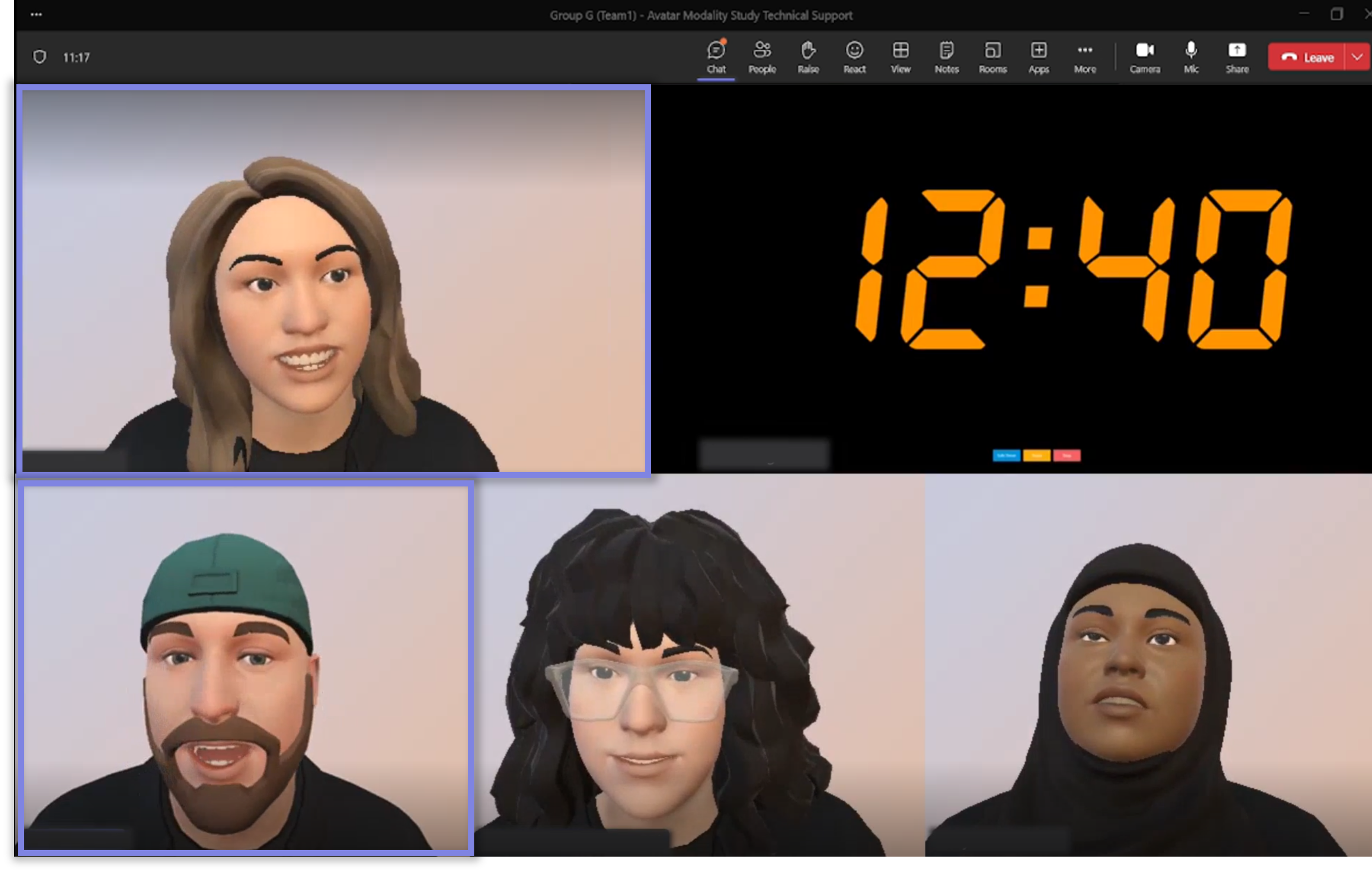}
  \Description{Microsoft Teams meeting window}
  \caption{A Microsoft Teams meeting window snapshot during task session.} 
  \label{fig:interface}
\end{figure} 

\autoref{fig:interface} shows an example during the actual experimental meeting, featuring a timer that counts down for decision-making tasks. The purple highlighted frames on the windows of the left two participants indicate the current speaker during ongoing discussions. This is a feature in Microsoft Teams.

\section{Quantitative Study}

\subsection{Demographics}

The following \autoref{tab:demographics} presents the demographic data collected from participants via the pre-session questionnaire, which provides a detailed demographic breakdown of the participants involved in the study, organized by gender and age group across the 16 formed groups.

\begin{table}
\centering
\resizebox{.7\textwidth}{!}{%
\begin{tblr}{
  hline{2} = {-}{},
}
Group\# & Female & Male & 18-24 & 25-34 & 35-44 & 45-54 & 55-64 & Total \\
1     & 3      & 1    & -     & 3     & -     & 1     & -     & 4     \\
2     & 3      & 1    & -     & 4     & -     & -     & -     & 4     \\
3     & 1      & 3    & -     & 1     & 1     & 3     & -     & 4     \\
4     & 1      & 3    & -     &       & 1     & 2     & 1     & 4     \\
5     & 3      & 1    & -     & 2     & 2     & -     & -     & 4     \\
6    & 1      & 3    & -     & 1     & 1     & 2     & -     & 4     \\
7    & 1      & 3    & -     & 1     & 2     & 1     & -     & 4     \\
8     & 1      & 4    & -     & -     & 4     & 1     & -     & 5     \\
9     & 2      & 3    & -     & 1     & 2     & 2     & -     & 5     \\
10     & 2      & 2    & -     & 2     & -     & 2     & -     & 4     \\
11     & 2      & 2    & 1     & 2     & 1     & -     & -     & 4     \\
12     & 0      & 4    & -     & -     & 2     & 2     & -     & 4     \\
13     & 0      & 4    & -     & -     & 2     & 2     & -     & 4     \\
14     & 2      & 2    & -     &       & 2     & 2     & -     & 4     \\
15     & 0      & 4    & -     & 1     & 2     & 1     & -     & 4     \\
16     & 2      & 3    & 1     & -     & 2     & 1     & -     & 5     \\
Total & 24     & 43   & 2     & 18    & 24    & 22    & 1     & 67    
\end{tblr}
}
\caption{Demographic breakdown of participants by gender and age group across 16 formed groups.}
\label{tab:demographics}
\end{table}

\subsection{Equation and Matlab function}

Test of Two Proportions equation is:
\begin{equation}
  \label{eq}
z = \frac{p1 - p2}{\sqrt{p(1-p)(\frac{1}{n1}+\frac{1}{n2})}}
\end{equation}

The Matlab function is:

\begin{lstlisting}[label={lst:label},
frame=single,
numbers=left,
style=Matlab-Pyglike]
function [p1,p2,z] = proportionsTest(r1,n1,r2,n2)
p1 = r1/n1;
p2 = r2/n2;
q1 = 1 - p1;
q2 = 1 - p2;
z = (p1-p2)/sqrt((p1*q1/n1) + (p2*q2/n2));
p1 = normcdf(-abs(z),0,1); % One tail P Value
p2 =2*p1; %two tail P Value
\end{lstlisting}

\newpage
\section{Qualitative Study}

\subsection{Quotes in Qualitative Analysis}
\autoref{tab:codingquote} provides sample sentences illustrating Meeting Satisfaction factors (\textit{effectiveness} and \textit{comfort}) coding references, as well as Avatar Satisfaction factors (\textit{preference}), derived from open-ended textual questions in the post-focus group questionnaire, used in thematic analysis.

\subsection{Additional Qualitative Sankey Flow Interpretation}

\begin{figure}[h]
  \centering
  \includegraphics[width=\linewidth]{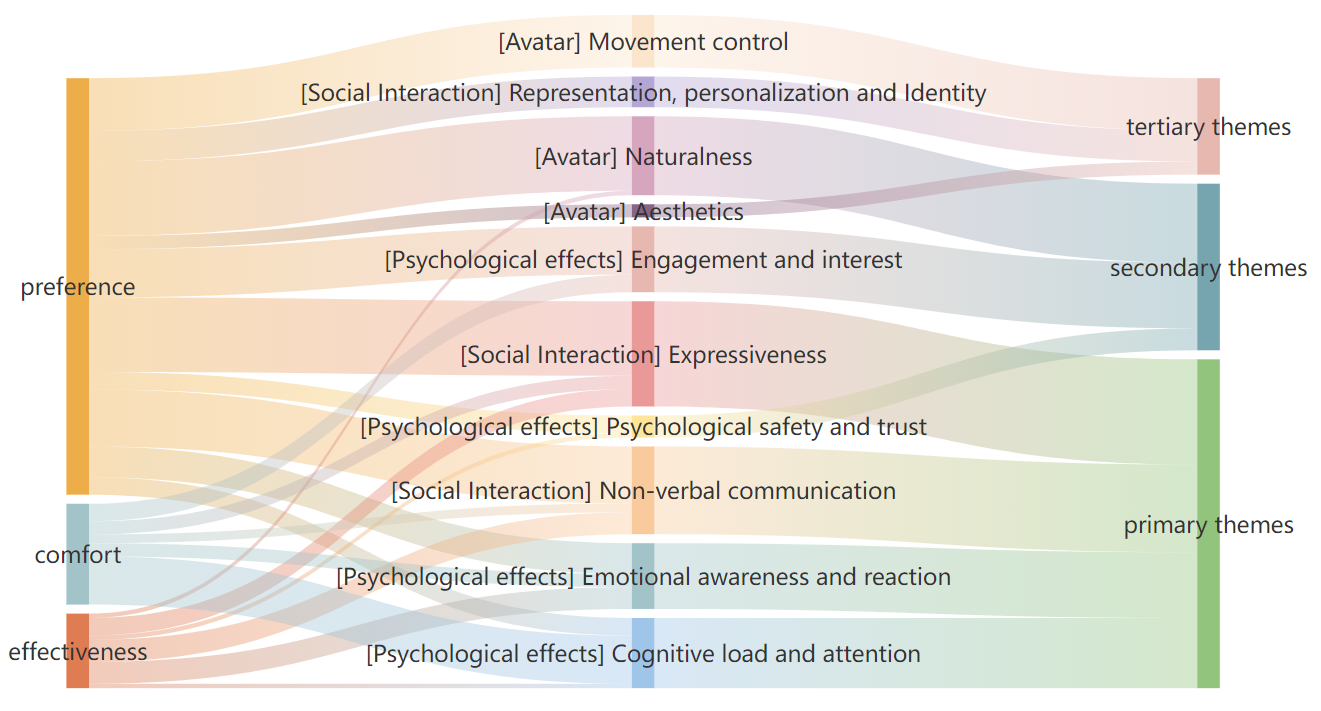}
    \Description{Sankey diagram}
    \caption{Sankey diagram illustrating the thematic flow recategorized into primary themes, secondary themes, and tertiary themes} 
    \label{fig:flow}
\end{figure}

This qualitative analysis flow is an extra interpretation through the use of a Sankey graph for \autoref{fig:flow} in the paper. This visualized graph facilitates both inductive (\textit{preference}) and deductive (\textit{effectiveness} and \textit{comfort}) thematic analysis flow for understanding the impact of themes of Avatar Satisfaction on Meeting Satisfaction. This approach allows the exploration and discovery of patterns, themes, and insights within qualitative themes. 

\begin{table}[h] 
\begin{small}
  \begin{center}
  \resizebox{1\textwidth}{!}{%
  \begin{tabular}{p{2cm}p{1.3cm}p{12cm}}
      \toprule
 \textbf{Theme name}   &\textbf{Factor}                &  \textbf{Sample sentences}  \\
    \toprule
    \multirow{2}{2cm}{1 - Naturalness} 
    &  \textit{preference} & \textit{“most life-like“, "most natural one because I can see the feedback from my movements”, “closest to real video stream while having movement”}\\
    & \textit{effectiveness} & \textit{“I like to see people’s expressions, understand what they think as well as what they say... and the [video driven] seems more natural to me”}\\
    \hline

    \hline
    {2 - Movement control} &  \textit{preference} & \textit{"I was constantly using [webcam-animated modality] to nod or shake my head to make things seem more natural“, "having my avatar move with my movements is critical vs having a static version or just mouth move - that felt distracting" , "I found [webcam-animated modality] to be really accurate to my face! I couldn't’t stop staring at it!"}\\
    \hline
    {3 - Aesthetic} &  \textit{preference} & \textit{“Face was less fat“, "My head was leaning forward in the other modalities"} \\

    \hline
    \multirow{3}{2cm}{4 - Expressiveness}  &  \textit{preference} & \textit{“I like the expressivity", "It is more interactive than the others", "I feel like it better represents me and enables me to better express myself with some facial expressions and movements"}\\
    & \textit{effectiveness} & \textit{"I felt a slight preference for webcam driven because I am someone who relies heavily on facial expressions to track the conversation"}\\
    & \textit{comfort} & {"The [webcam-animated] was the most tiring because it encouraged me to move more" , "The [picture modality] was more tiring because it was only picture therefore requiring a little more work on my end to effectively express myself"}\\
    \hline
    \multirow{3}{2cm}{5 - Non-verbal Communication}
    &  \textit{preference} & \textit{"We know when people are talking [in webcam-animated modality modality]", "The [picture modality] was somewhat exhausting, as it presented difficulties in discerning the appropriate moments to contribute to the discussion“,  "[In picture and audio-animated modalities] cannot tell whether people are ready to start or finish talking“, “I was able to convey emotions and reactions using non-verbal communication (head nodding, smiling, etc) without having to rely on a video call"}\\
    & \textit{effectiveness} & \textit{“It was helpful to see smiles and head nods as feedback to my statements"}\\
    & Comfort & \textit{[picture and audio-animated modality] was less tiring for my body, but more tiring for my brain because I had to think about what people might be feeling or taking away from the meeting without any visual confirmation. Also, it was harder to know when to jump into the conversation (when no one else was talking)}\\
    \hline
    {6 - Representation, Personalization, and Identity}
    &  \textit{preference} & \textit{"the [webcam-animated modality] felt like it allowed my personality to come across more in the discussion", "the [webcam-animated modality] was the most expressive, and made me feel most "like myself" in a meeting“, "because it’s important to try new experience not losing the identity so I chose the [webcam-animated modality]"}\\
    \hline
    \multirow{2}{2cm}{7 - Engagement and Interest}
    &  \textit{preference} & \textit{"I like the [webcam-animated modality] animations. They remind me that I’m speaking with real people in real time. “ , “[The webcam-animated modality is] the most engaging and lifelike." "This was fun!“, "the [webcam-animated modality] seemed like the most interesting one because it is dynamic."}\\
    & Comfort & \textit{I don’t that any were more tiring, but [audio-animated and picture modality] were both boring. [picture modality], in particular. If it’s just a picture, I’d rather see their profile photo"}\\  
    \hline
    \multirow{3}{2cm}{8 - Emotional Awareness and Reaction}
    &  \textit{preference} & \textit{"I appreciated that I was able to convey emotions and reactions using non-verbal communication by [webcam-animated modality]," "That [webcam-animated modality] makes me feel more comfortable and confident while I can see my peers’ emotions"}\\
    & \textit{effectiveness} & \textit{“the topics with picture, and with broader options were more difficult, as on scenario 3, as we could not see the people reactions”}\\
    & Comfort & {"the pictures were more tiring as we could not see the reactions of the people"}\\
    \hline 
    \multirow{3}{2cm}{9 - Cognitive Load and Attention}
    &  \textit{preference} & \textit{“The [webcam-animated modality] encouraged me to stay focused on the task and discussion“, “ I like the expressively and lower cognitive burden of trying to deduce other people’s states / make my own clear by voice only}\\
    & \textit{comfort} & \textit{"I found the [picture modality] where folks were frozen to be more tiring. I had to force myself to pay attention to the conversation".}\\
    & \textit{effectiveness} & \textit{“Video was more natural for communication and had less cognitive load, but I became more self-aware and self-conscious about what my avatar was doing at any given moment (was my mouth open, weird face angle, etc)“, "Session 2 is more tiring as it makes you more conscious of what you are doing"}\\
    \hline
    \multirow{2}{2cm}{10 - Psychological Safety and Trust}
    &  \textit{preference} & \textit{"I chose the [webcam-animated modality] because it feels more natural and trustworthy“,  "I appeared as the [webcam-animated modality] because it felt expressive but didn’t invade my privacy like a full-on camera would."}\\
    & \textit{effectiveness} &\textit{"[webcam-animated modality] is for sure the more complete to create a trust with other people joining the call"}\\
    \bottomrule
  \end{tabular}}
  \end{center}
  \end{small}
  \caption{Quotes associated with the 10 identified themes across the three focus group questions (
  $\textit{Q}_{\textit{\small{AS-pref-text}}}$ ,$\textit{Q}_{\textit{\small{MS-eff-text}}}$ and $\textit{Q}_{\textit{\small{MS-comf-text}}}$ 
  )}
  \label{tab:codingquote}
\end{table} 
\FloatBarrier

\newpage
\subsection{Thematic Analysis Coding Process: Branching Diagram}

These coding process diagrams that are generated by Nvivo12, specifically pertain to Avatar and Meeting Satisfaction. The branching directions are provided to demonstrate a deeper understanding of initial codes and 10 themes in this paper which are analysed from both deductive coding and inductive coding procedures. 

\begin{figure}[h]
  \centering
  \includegraphics[width=\linewidth]{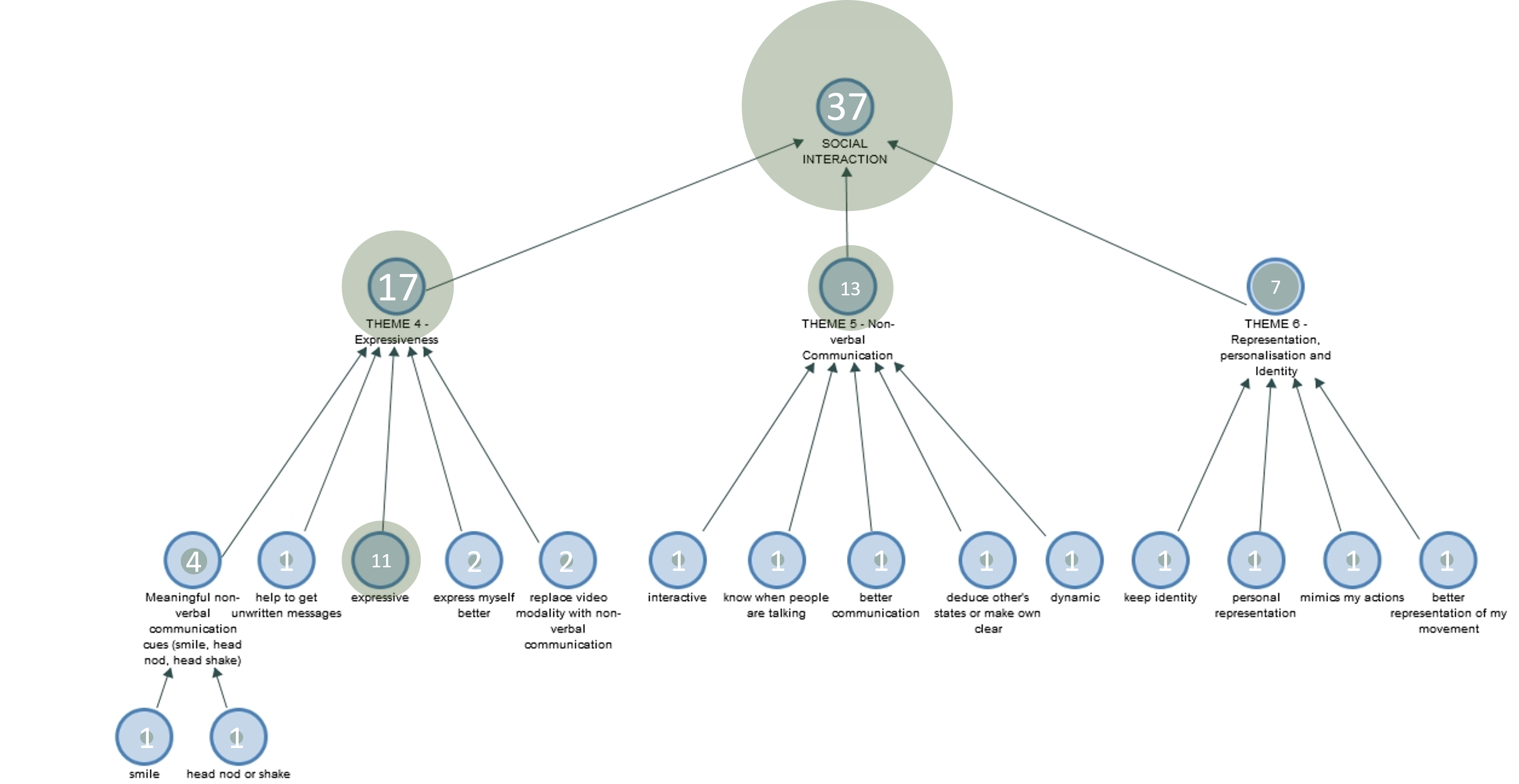}
    \caption{Inductive coding branch for Avatar Satisfaction Factor \textit{preference} ($\textit{Q}_{\textit{\small{AS-pref-text}}}$): Expressiveness, Non-verbal communication and Representation, personalision and Identity in category Social Interaction} 
    \label{fig:mpf2coding}
    \Description{Inductive coding branch}
\end{figure}

\begin{figure}[h]
  \centering 
  \includegraphics[width=\linewidth]{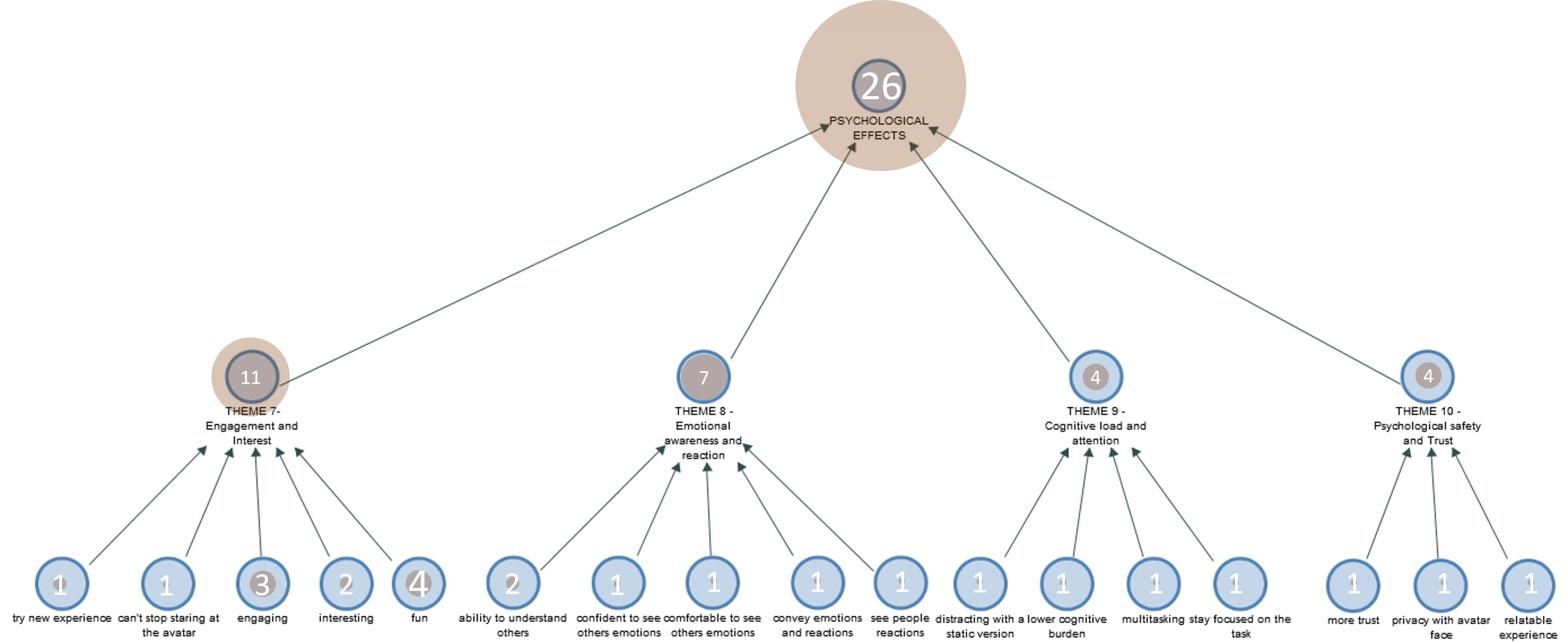}
    \caption{Inductive coding branch for Avatar Satisfaction Factor \textit{preference} ($\textit{Q}_{\textit{\small{AS-pref-text}}}$): Engagement and interest, Emotional awareness and reaction, Cognitive load and attention and Psychological safety and trust in category Psychological Effects} 
    \label{fig:mpf3coding}
        \Description{Inductive coding branch}
\end{figure}

\textbf{Inductive Coding Process} Bottom-up process in \autoref{fig:mpf1coding}, \autoref{fig:mpf2coding}, \autoref{fig:mpf3coding}, shows three categories (Avatar, Social Interaction and Psychological Effects) of Avatar Satisfaction factor \textit{preferences} ($\textit{Q}_{\textit{\small{AS-pref-text}}}$) with the reference count of each theme:

\begin{figure}[h] 
  \centering
  \includegraphics[width=\linewidth]{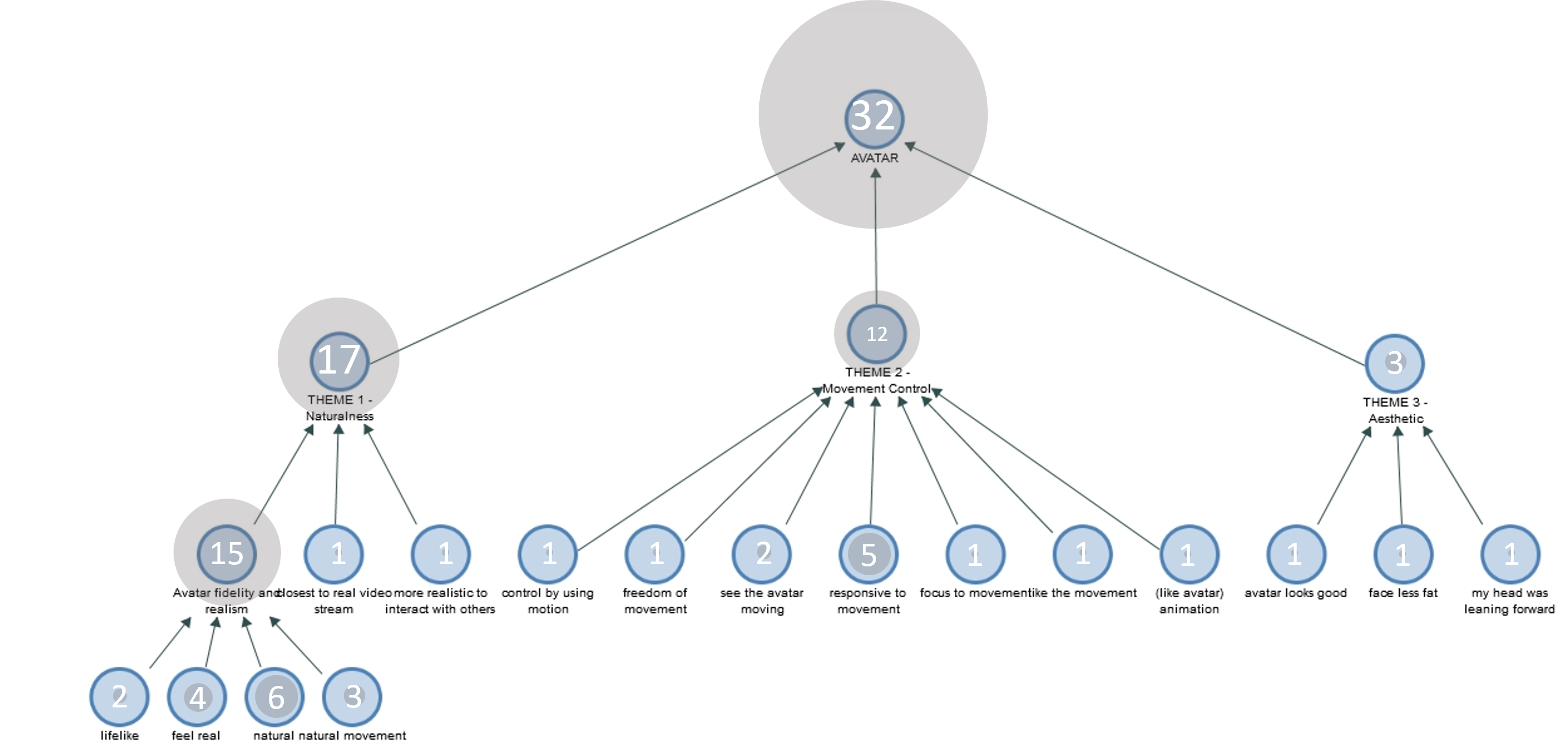}
    \caption{Inductive coding branch for Avatar Satisfaction Factor \textit{preference} ($\textit{Q}_{\textit{\small{AS-pref-text}}}$): Naturalness, Movement control and Aesthetic in category Avatar} 
    \label{fig:mpf1coding}
        \Description{Inductive coding branch}
\end{figure} 

\begin{figure}[h]
  \centering
  \includegraphics[width=0.7\linewidth]{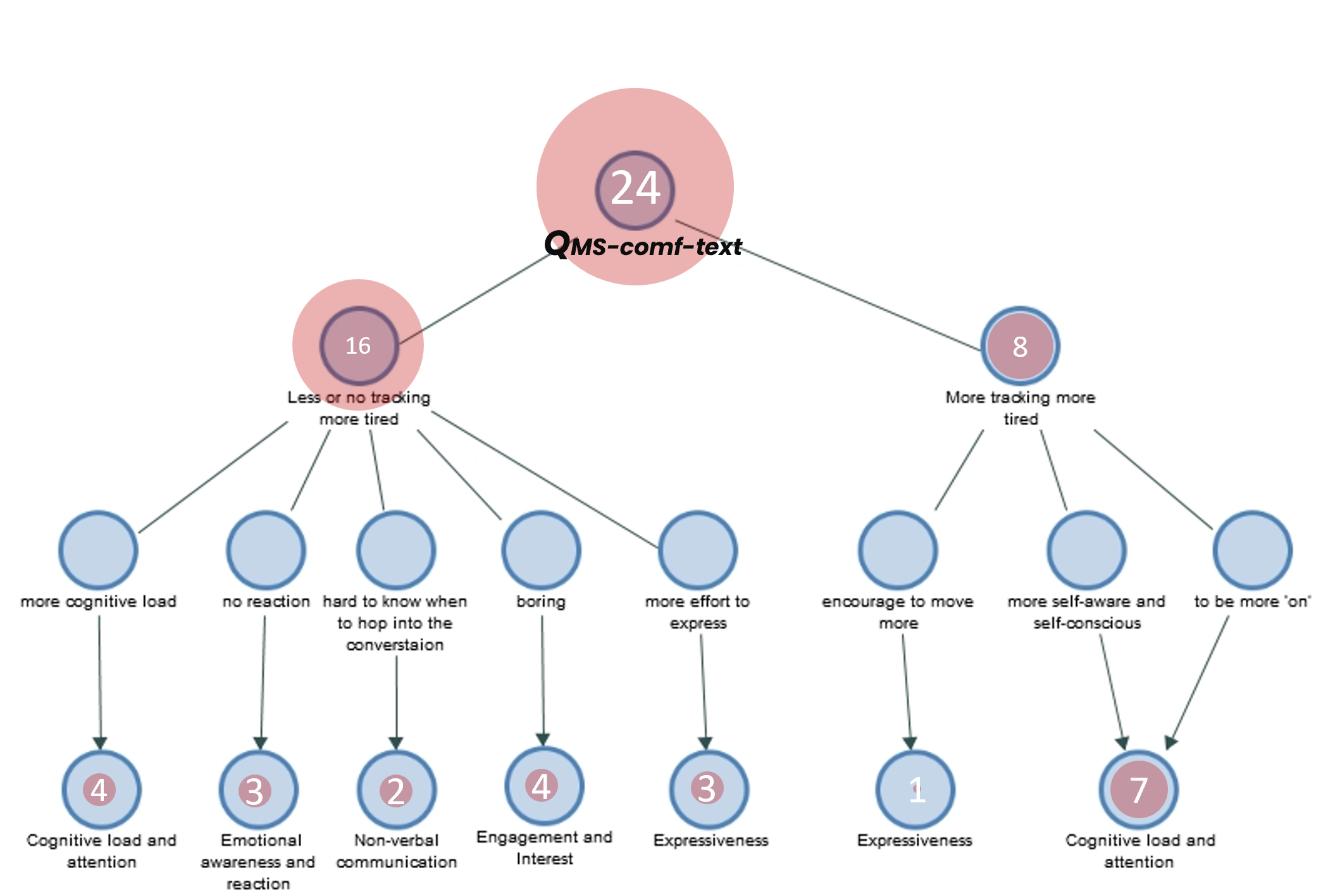}
  \caption{Deductive coding branch for Meeting Satisfaction Factor \textit{comfort} ($\textit{Q}_{\textit{\small{MS-comf-text}}}$) includes 5 themes: Expressiveness, Non-verbal communication and Representation, Emotional awareness and reaction, Cognitive load and attention, Engagement and interest} 
  \label{fig:msf2coding}
        \Description{Deductive coding branch}
\end{figure}

\newpage
\textbf{Deductive Coding Process} Top-down process in \autoref{fig:msf1coding} and \autoref{fig:msf2coding} for two Meeting Satisfaction factors \textit{effectiveness} ($\textit{Q}_{\textit{\small{MS-eff-text}}}$) and \textit{comfort} ($\textit{Q}_{\textit{\small{MS-comf-text}}}$) with the reference count of each theme:

\begin{figure}[h]
  \centering  
  \includegraphics[width=0.8\linewidth]{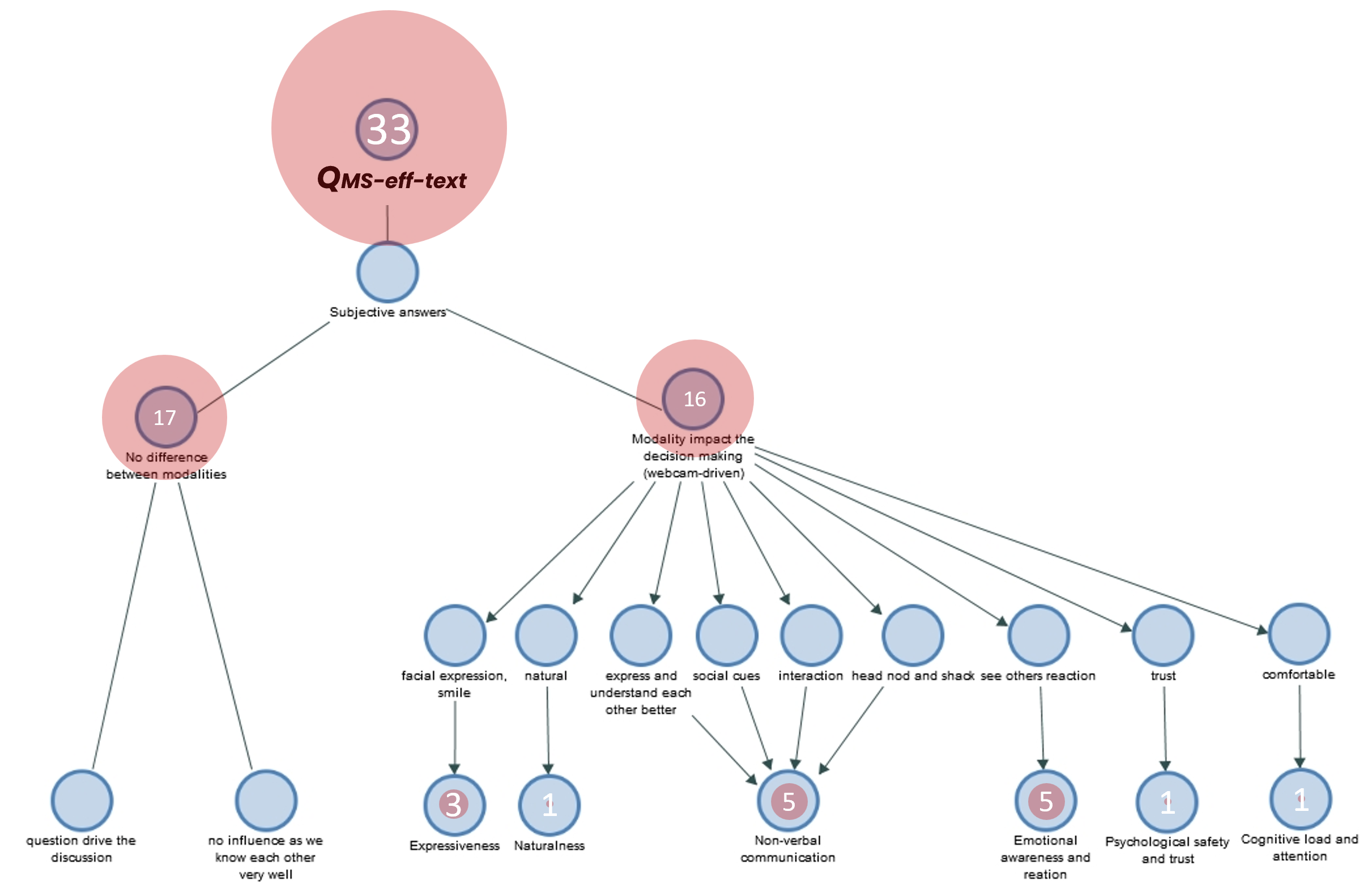}
  \caption{Deductive coding branch for Meeting Satisfaction Factor \textit{effectiveness} ($\textit{Q}_{\textit{\small{MS-eff-text}}}$) includes 6 themes: Expressiveness, Non-verbal communication and Representation, Emotional awareness and reaction, Cognitive load and attention, Naturalness, Psychological safety and trust} 
  \label{fig:msf1coding}
      \Description{Deductive coding branch}
\end{figure}

